\documentclass[preprint,
aps,amssymb,nofootinbib,showpacs,
tightenlines,showkeys,secnumarabic]{revtex4}
\usepackage{graphicx}
\usepackage[section]{placeins}
\newcommand{\be}{\begin{equation}}
\newcommand{\ee}{\end{equation}}
\newcommand{\ba}{\begin{eqnarray}}
\newcommand{\ea}{\end{eqnarray}}
\newcommand{\nn}{\nonumber}

\newcommand{\rt}{\langle n\rangle}
\newcommand{\vrt}{\langle v\rangle}
\newcommand{\brt}{\begin{ruledtabular}}
\newcommand{\ert}{\end{ruledtabular}}
\newcommand{\hl}{\hline}

\begin{document}

\title{Inverted regions induced by geometric constraints on a 
classical encounter-controlled binary reaction}

\author{E. Abad}
\thanks{Corresponding author}
\email{eabad@ulb.ac.be}
\affiliation{Center for Nonlinear Phenomena and Complex Systems,
Universit\'e Libre de Bruxelles
C.P. 231, 1050 Bruxelles, Belgium}

\author{John J. Kozak}
\affiliation{Beckman Institute, California Institute of Technology,
Pasadena CA 91125-7400 \footnote{Permanent address: 
DePaul University, 243 South Wabash Avenue, Chicago IL 60604-2301}}



\begin{abstract}

The efficiency of an encounter-controlled reaction between two
independently-mobile reactants on a
lattice is characterized by the mean number $\rt$ of steps to reaction. 
The two reactants are
distinguished by their mass with the "light" walker performing a jump to
a nearest-neighbor site in each
time step, while the "heavy" walker hops only with a probability $p$;  we
associate $p$ with the
"temperature" of the system. To account for geometric exclusion effects
in the reactive event, two reaction
channels are specified for the walkers; irreversible reaction occurs
either in a nearest-neighbor collision,
or when the two reactants attempt to occupy the same site.  Lattices
subject to periodic and to confining
boundary conditions are considered. For periodic lattices, depending on
the initial state, the reaction
time either falls off monotonically with $p$ or displays a local minimum
with respect to $p$; occurrence of the latter signals a regime where the
efficiency of the reaction effectively decreases with increasing
temperature. Such behavior can also occur when one averages over all initial
conditions, but can disappear if the jump
probability of the light walker falls below a characteristic threshold
value.  Even more robust behavior can
occur on lattices subject to confining boundary conditions.  Depending on
the initial conditions, the reaction
time as a function of $p$ may increase monotonically, decrease
monotonically, display a single maximum
or even a maximum and minimum;  in the latter case, one can identify
distinct regimes where the above-noted
inversion in reaction efficiency can occur. We document both numerically
and theoretically that these
inversion regions are a consequence of a strictly classical interplay
between excluded volume effects implicit in the specification of the two
reaction channels, and the system's dimensionality and spatial extent.
Our results highlight situations where the description of an
encounter-controlled reactive event cannot
be described by a single, effective diffusion coefficient.  We also
distinguish between the inversion
region identified here and the Marcus inverted region which arises in
electron transfer reactions.
 
\end{abstract}

\maketitle
\section{Introduction}

Understanding the interplay between diffusion-controlled dynamics and
the geometrical constraints
imposed by the substrate or topology of the host medium is a problem of
seminal importance today. For example, with recent advances in 
nanotechnology, one can now
visualize and manipulate individual
atoms and molecules, and design specific geometries on atomic length
scales \cite{kar1,kar2,kar3}. The geometrical
characteristics of the latter (i.e., size, dimensionality and boundary
conditions) are found to influence
the efficiency of diffusion-controlled reactive processes and trapping
events governed by specific interactions.

After the seminal papers by Ovchinnikov and 
Zeldovich \cite{ovch} and Toussaint and Wilczek \cite{touss}
documenting an anomalous decrease in the efficiency of the 
two-species annihilation reaction in low dimensions due to 
reactant segregation effects, other types of encounter-controlled 
processes between unlike species have also received much attention
in the literature. In particular, the so-called trapping problem and 
target problem have been extensively studied \cite{weiss,redn}. 
Such processes are ubiquitous in nature and the underlying
theory finds wide application in domains such as trapping of 
charge carriers \cite{seid}, 
exciton annihilation and spin relaxation processes
\cite{racz}, quenching of delocalized excitations,
recombination and coagulation 
processes \cite{rice}, predator-prey models \cite{red4,redn} etc. 
In the simplest version of the trapping problem, diffusing $A$ particles
are annihilated upon encounter with immobile traps $T$ following the
scheme $A+T\rightarrow T$. The target problem may be thought of as dual
to the trapping problem \cite{red0,blum}, 
i.e., here randomly moving $A$ particles are
the ones which destroy immobile targets $T$ upon encounter
according to the reaction $A+T\rightarrow A$. The case where
both $A$ particles and traps diffuse with different diffusivities
is significantly harder to deal with and few analytic solutions 
are known \cite{brams, szab, bray1, osh1, mor1, mor2}. 
With a few exceptions, most of the previous work on such variants of 
the trapping and the target problem has been focused on the computation 
of the survival probability $S(t)$ for the annihilated species
and the long-time asymptotics of this property
as a measure of the reaction's efficiency.       

On the other hand, a number of papers \cite{nickoz1, jon1, jon2} 
have taken up the problem of characterizing the efficiency of 
two-particle, encounter-controlled reactions 
on a lattice. The type of reaction considered in those references 
can also be interpreted 
as a space- and time-discrete version of the trapping problem 
or the target problem as defined above. In contrast to previous work, 
the quantity used to gauge the efficiency of the process was the 
initial-condition-averaged mean number of time 
steps to reaction $\langle n\rangle$, whose inverse 
is a measure of the time scale of the reaction and hence the efficiency
of the reaction-diffusion event; in particular, in the
large lattice limit the inverse converges (in magnitude) to the smallest
eigenvalue of the associated master equation \footnote{The counterpart 
$\langle t \rangle$ of $\rt$ in the diffusive limit is connected with 
the survival
probability via the relation $\langle t\rangle=\int_0^{\infty} S(t)\,dt$,
(see e.g. \cite{weiss}, pages 173-174)}. 
Basically two types of situations were considered: a) 
an asynchronous scenario in which a "light" walker (also termed 
``walker 1'' hereafter) 
performs nearest-neighbor jumps at each time step 
while a "heavy" walker (walker 2) remains stationary at a given site,
thereby playing the role of an immobile trap and b)
a situation where both walkers synchronously perform 
symmetric nearest-neigbor jumps on the lattice. Synchronous dynamics 
turns out to be more efficient in general, but significant 
exceptions are found for sufficiently small lattices. 

If one assumes, for instance in the framework of
Arrhenius theory, that the mobility of the heavy walker grows with
increasing system temperature, one realizes that such exceptions
are quite remarkable. In classical diffusion theory, as formulated by
von Smoluchowski and Einstein, the diffusion coefficient 
is a monotonically increasing function of the temperature. Thus, in a
 continuum approximation where one identifies an effective 
diffusion coefficient the efficiency of an encounter-controlled
reaction should increase with increasing temperature. Deviations from this
expectation emphasize the fact that some properties (particularly for 
small systems) are not correctly described by an approach valid
for large systems in the thermodynamic
limit. It is precisely these properties that are of special importance 
in applications of nanotechnology. 

To understand in a more detailed way the role of temperature in influencing
the efficiency of diffusion-controlled events in small systems, the 
approach taken in \cite{nickoz1, jon1, jon2} was generalized
in \cite{eabad} as follows: a jump probability $p$ of walker 
was defined such that $p=0$ corresponded to purely asynchonous behavior 
and $p=1$ corresponded to purely synchronous behavior of the two diffusing
particles \footnote{Note that, even though we have chosen 
to interpret the difference in the mobility of both walkers for $p<1$ 
as a difference
in their masses, other interpretations are also legitimate. In 
nonequilibrium system one could e.g. also speak of a 
difference in their internal energies.}. Interpreting the parameter $p$
as a measure of the ``system temperature'', 
one would intuitively anticipate that an increase in temperature should 
lead to an increase in the efficiency of reaction. For the model studied in 
\cite{eabad} and this paper, this
expectation was confirmed for certain initial conditions. 
However, already in 1D, it was shown that
for properly chosen initial conditions the model leads to the onset of
a parameter region where $\langle n\rangle$ increases monotonically
with $p$, implying that the efficiency of the
reaction decreases with increasing temperature.

In what follows, we shall term this region ``inverted'' by analogy
with the well-known decrease of the reaction rate 
with increasing temperature predicted by Marcus for
electron transfer reactions \cite{mar1,mar2,mar3} in certain
regimes of parameter space (reaction coordinate, reorganization energy)
and later confirmed
experimentally \cite{closs,gray1,gray2}. However, we emphasize that our 
use of this terminology is purely
symbolic, since in the cases with which we shall be dealing here the 
anomalous decrease in efficiency is due to a stricly classical 
interplay between
the system geometry and an ``interaction zone'' (to be specified
in Sec. \ref{Sec2a}) in binary collisions,
rather than a quantum-mechanical
analysis of curve crossing between reactant
  and product states (where Marcus showed that there are regions in
which the reaction probability can
  actually decrease with increase in "driving force" of the reaction).
As we shall show, when the size of the "interaction zone" of the two 
reactants becomes of the same order of magnitude as the system size,
anomalous behavior is found.

To explore the relevance of
our work to possible experimental realizations,
it will be necessary to study in detail the extent to which our
"inversion region" is robust with respect to system
dimensionality and/or boundary conditions. For example, the difference
equation approach developed in \cite{eabad}
for 1D periodic lattices cannot be extended to treat the case of higher
dimensions or non-translationally
invariant lattices. Accordingly, we apply here the theory of finite
Markov processes  to investigate the
previously-noted, more general cases in detail.  Further, we shall
consider cases where the jump probability
of the "light walker" is restricted to values less than one (and will
find that the "inversion region" disappears
when the jump probability falls below a threshold value).  Finally, we
shall propose (several) rationale to explain
the quantitative behavior uncovered in our study.

The plan of the paper is then the following: 
Sec. \ref{Sec2a} gives a general definition
of the model and shows how in the framework of Markov chain theory 
the mean reaction time and its variance can be extracted from the
so-called fundamental matrix (the size of this matrix can be greatly
reduced by proper use of the system symmetry).
In Sec. \ref{Sec2}, we deal with the case of a 1D lattice, thereby
comparing the results for the periodic and the confining case. This is
done both
for specific initial conditions and at a coarse grained level (average
over a homogeneous set of initial states).  
Sec. \ref{Sec4} treats the case of a 2D lattice, both for periodic and
confining boundary conditions. Sec. \ref{Sec5} discusses the case 
where the light walker is allowed to jump
 with probability less than one at each time step.  
Sec. \ref{Sec6} gives a physical interpretation of the main results. 
Finally, Sec. \ref{Sec7} summarizes the
conclusions and discusses possible extensions of the model.  
The appendices A and B contain tables with detailed analytic 
expressions for the reaction time in the general case treated in
Sec. \ref{Sec5}. 
 
\section{General framework} 
\label{Sec2a}

\subsection{Definition of the model and use of symmetry}
\label{defsym}

Consider two reactants (random walkers labeled, respectively, 1 and 2) 
performing a time discrete 
random walk on an $N$-site lattice with dimensionality
and boundary conditions to be specified later. At each time step, walker 1  
performs a symmetric nearest-neighbor jump with probability one, 
while walker 2 either 
performs a symmetric nearest-neighbor jump with probability
$p$ (synchronous step) or remains at the same site (asynchronous
step). Any time the walkers jump on the same site or exchange
positions by crossing each other, the reaction takes place instantaneously,
and the process is terminated. Our main goal will be to clarify how
the $p$-dependance of the mean number of time steps to reaction 
changes with the boundary conditions and the dimensionality of the
system and to ellucidate the role of the initial conditions in this 
context.  

\begin{figure}[htbp]
\includegraphics[width=8cm,height=.75cm]{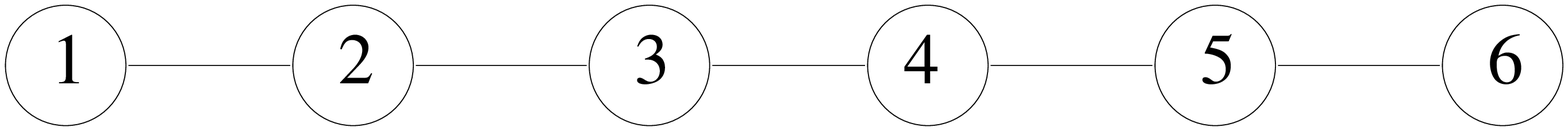}
\hspace{2.5cm}
\includegraphics[width=4cm,height=4cm]{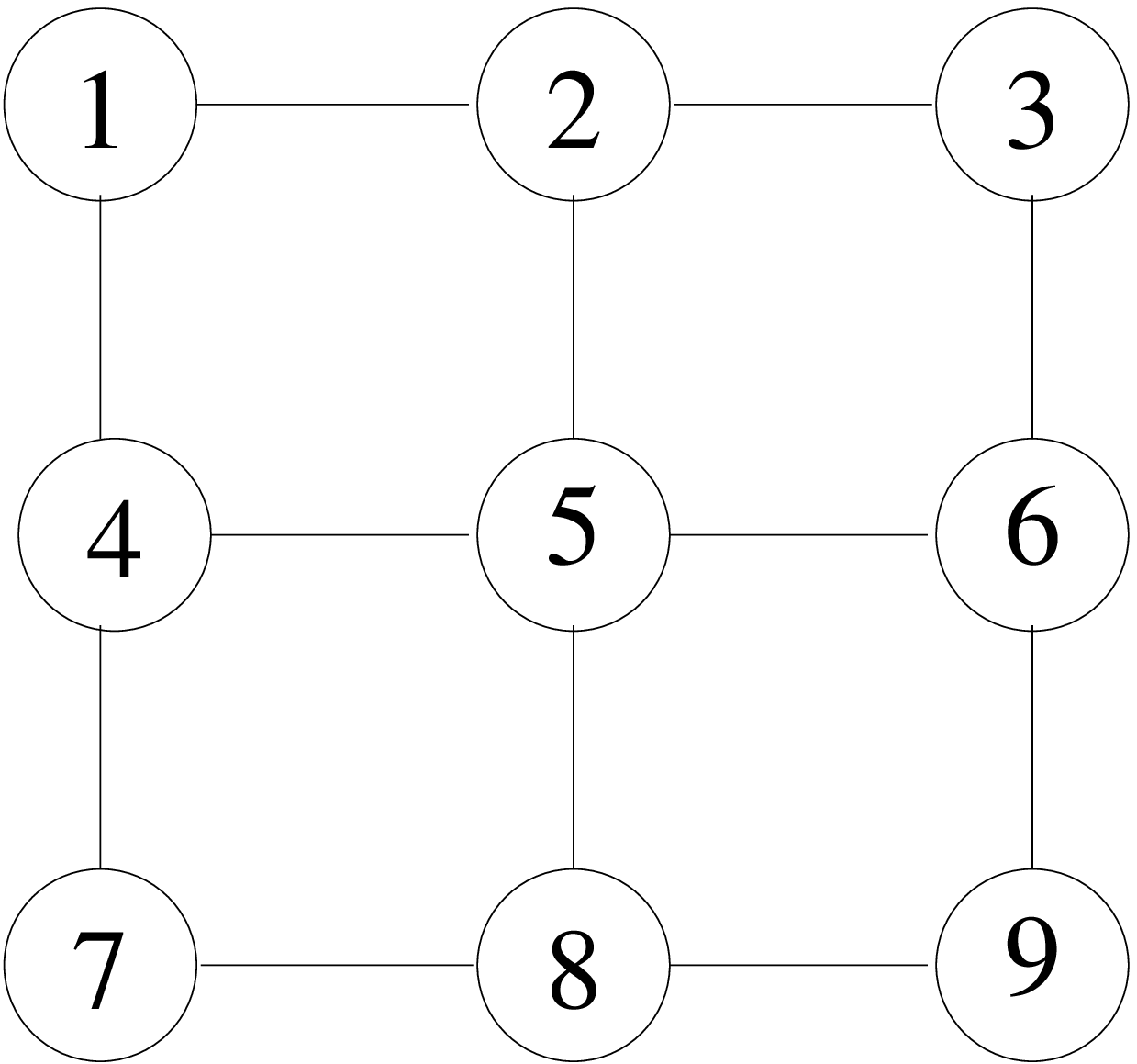}
\caption{\it \label{fig0} Integer coordinates associated with
each site for a 1D lattice with $N=6$ (left) and a 2D lattice 
with $N=9$ (right).}
\end{figure}

The stochastic motion of the two walkers in the above model can 
be considered as an absorbing Markov chain. In order to characterize
the transient states of the chain let us first label the lattice sites
from 1 to $N$.
In the case of a 1D lattice this can e.g. be done from left to right 
(see Fig. \ref{fig0}, left),
while in the case of a 2D lattice one can number the lattice sites, say, 
from left to right and from top to bottom
(see Fig. \ref{fig0}, right). 
We shall consider two types of boundary conditions, 
i.e. periodic and confining. By confining boundaries (also sometimes
termed 'reflecting' in the literature)
we mean that whenever either of the two walkers attempts to 'exit' the 
lattice from a boundary site under the dynamics prescribed above, 
it is reset to the same position.

Each transient state can thus be expressed by a pair of positive 
integers $x_1$ and $x_2$ representing 
the instantaneous position of walker 1 and walker 2 in terms of the
above lattice site coordinates. Let us hereafter denote by 
$x_1^0$ and $x_2^0$ the initial values of 
these coordinates. Thus, there are $N(N-1)$ possible pairs
of ``non-reactive'' initial two-particle configurations $(x_1^0,x_2^0)$.
Each of these configurations has an associated 
local reaction time $\rt_{(x_1^0,x_2^0)}$, whereby the angular
brackets denote the average over a uniform ensemble of joint random
walk realizations. In addition to these quantities, one can construct
an expression for the global reaction time $\rt$, i.e., the average over 
a uniform set of initial conditions $(x_1^0,x_2^0)$: 

\be
\rt=\frac{1}{N(N-1)}\sum_{x_2^0=1 
\atop x_2^0\neq x_1^0}^N \sum_{x_1^0=1}^N \rt_{(x_1^0,x_2^0)}.
\ee 

In the present work we shall make an extensive use of symmetry to simplify 
the calculation of $\rt$. Let us e.g. consider the case of a 1D lattice
with $N=4$. The set of initial configurations obtained by 
assigning different values to $x_1^0$ and $x_2^0$ can be represented 
pictorially as follows:

\[
\begin{array}{c}
\mbox{State (2,1)}:\qquad 2-1-0-0 \\
\mbox{State (3,1)}:\qquad 2-0-1-0 \\
\mbox{State (4,1)}:\qquad 2-0-0-1 \\
\mbox{State (1,2)}:\qquad 1-2-0-0 \\
\mbox{State (3,2)}:\qquad 0-2-1-0 \\
\mbox{State (4,2)}:\qquad 0-2-0-1 \\
\mbox{State (1,3)}:\qquad 1-0-2-0 \\
\mbox{State (2,3)}:\qquad 0-1-2-0 \\
\mbox{State (4,3)}:\qquad 0-0-2-1 \\
\mbox{State (1,4)}:\qquad 1-0-0-2 \\
\mbox{State (2,4)}:\qquad 0-1-0-2 \\
\mbox{State (3,4)}:\qquad 0-0-1-2 
\end{array}
\]

,where the ``1(2)'' represent sites occupied by walker 1(2) and the 
``$0$'' represent empty lattice sites. 

In the periodic case, the state of the system is fully 
characterized by the distance 
$d=\mbox{min}(|x_1-x_2|, N-|x_1-x_2|)$ between the walkers. 
Thus, the above $12$ states can be lumped into the first $2$ states
(2,1) and (3,1), respectively numbered
as State 1 and State 2 in the short-hand notation to be used in the following. 

Let us now consider the case of a confining lattice. 
In addition to the first two states, 
one has 4 additional symmetry-distinct states, 
e.g. (4,1), (3,2), (4,2) and (4,3). In our short-hand notation we thus have:

\[
\begin{array}{c}
\mbox{State 1}:\qquad 2-1-0-0 \\
\mbox{State 2}:\qquad 2-0-1-0 \\
\mbox{State 3}:\qquad 2-0-0-1 \\
\mbox{State 4}:\qquad 0-2-1-0 \\
\mbox{State 5}:\qquad 0-2-0-1 \\  
\mbox{State 6}:\qquad 0-0-2-1 
\end{array}
\]
This numbering procedure for the symmetry-distinct states can now 
straightforwardly be extended to 1D confining lattices with arbitrary 
$N$, i.e. by first setting $x_2=1$ and 
letting $x_1$ vary from 2 to $N$, then increasing $x_2$ by one unit 
and letting $x_1$ vary from $3$ to $N$, and so on. One thus obtains 
$N(N-1)/2$ states characterized
by $x_1>x_2$. In particular, the states for the cases $N=5$ and $N=6$
to be dealt with in \ref{cobc} respectively read

\[
\begin{array}{c}
\mbox{State 1}:\qquad 2-1-0-0-0 \\
\mbox{State 2}:\qquad 2-0-1-0-0 \\
\mbox{State 3}:\qquad 2-0-0-1-0 \\
\mbox{State 4}:\qquad 2-0-0-0-1 \\
\mbox{State 5}:\qquad 0-2-1-0-0 \\  
\mbox{State 6}:\qquad 0-2-0-1-0 \\ 
\mbox{State 7}:\qquad 0-2-0-0-1 \\
\mbox{State 8}:\qquad 0-0-2-1-0 \\  
\mbox{State 9}:\qquad 0-0-2-0-1 \\ 
\mbox{\hspace{-.35cm} State 10}:\qquad 0-0-0-2-1 
\end{array}
\]

and

\[
\begin{array}{c}
\mbox{State 1}:\qquad 2-1-0-0-0-0 \\
\mbox{State 2}:\qquad 2-0-1-0-0-0 \\
\mbox{State 3}:\qquad 2-0-0-1-0-0 \\
\mbox{State 4}:\qquad 2-0-0-0-1-0 \\
\mbox{State 5}:\qquad 2-0-0-0-0-1 \\
\mbox{State 6}:\qquad 0-2-1-0-0-0 \\ 
\mbox{State 7}:\qquad 0-2-0-1-0-0 \\
\mbox{State 8}:\qquad 0-2-0-0-1-0 \\  
\mbox{State 9}:\qquad 0-2-0-0-0-1 \\ 
\mbox{\hspace{-.35cm} State 10}:\qquad 0-0-2-1-0-0 \\
\mbox{\hspace{-.35cm} State 11}:\qquad 0-0-2-0-1-0 \\
\mbox{\hspace{-.35cm} State 12}:\qquad 0-0-2-0-0-1 \\
\mbox{\hspace{-.35cm} State 13}:\qquad 0-0-0-2-1-0 \\
\mbox{\hspace{-.35cm} State 14}:\qquad 0-0-0-2-0-1 \\
\mbox{\hspace{-.35cm} State 15}:\qquad 0-0-0-0-2-1 
\end{array}
\]

In the case of a periodic lattice, it is sufficient to 
consider the first
$[N/2]$ states, where $[\cdot]$ is the floor function 
(largest integer smaller than
the argument). In this case, the state number in the short-hand 
notation is identical with the interparticle distance $d$. 

In order to compute the global reaction time $\rt$ from the local
ones, one constructs

\be
\label{elsum}
\langle n\rangle=\sum_{i=1}^{N_s} w_i \langle n\rangle_i, 
\ee

where $N_s$ is the number of symmetry distinct states (respectively equal
to $[N/2]$ and $N(N-1)/2$ for a periodic and for a confining 1D lattice), 
$\langle n\rangle_i$ is the reaction time when both walkers start
from a given symmetry-distinct initial 
state $i$, and $w_i$ are appropiately chosen weights. The latter are easily
calculated, e.g. for the periodic 1D lattice one has 

\be
\label{we1}
w_i=\frac{1}{[N/2]} 
\ee
for odd values of $N$ and 

\begin{eqnarray}
w_i&=&\frac{2}{N-1}, \quad i=1,\ldots,[N/2]-1,\nonumber\\
\label{we2}
w_{[N/2]}&=&\frac{1}{N-1} 
\end{eqnarray} 
for even values of $N$. 
These expressions reflect the fact that for even lattices the statistical
weight of the initial condition with maximum distance [N/2] 
between the walkers is 
half the weight of the other initial states (which appear twice when
taking the average over a homogeneous ensemble of initial conditions), 
while in the odd lattice case all initial states have the same weight.

\subsection{Markovian approach}
\label{marap}

In the framework of Markov chain theory the mean number 
of time steps $\rt_i$ before termination of a random walk starting
from state $i$ is given by the elements of the 
so-called fundamental matrix. This matrix reads
  
\be
\label{invfor}
\overline{\overline{\bf N}}\equiv
\sum_{k=0}^{\infty} \overline{\overline{\bf Q}}^{\,\, k}=
(\overline{\overline{\bf I}} -
\overline{\overline{\bf Q}})^{-1},
\ee
where $\overline{\overline{\bf Q}}$ is the matrix whose elements
$q_{nm}$ are the transition probabilities between two transient
(non-reactive) states $n$ and $m$. The matrices 
$\overline{\overline{\bf Q}}$ and 
$\overline{\overline{\bf N}}$ thus 
have the dimension $N_s\times N_s$. Denoting by $n_{nm}$ the elements
of the fundamental matrix, one then has \cite{snell}

\be
\label{locrt}
\rt_i=\sum_{j=1}^{N_s}n_{ij}.
\ee

For small lattices the analytical computation of the transition matrix 
$\overline{\overline{\bf Q}}$ in terms of the 
jump probability $p$ as well as the 
subsequent derivation of the fundamental matrix 
is straightforward. The local and
the global reaction times 
are obtained from the matrix elements   
via eqs. (\ref{locrt}) and (\ref{elsum}). The variance 
$\langle v\rangle_i\equiv 
\langle n^2\rangle_i-\langle n\rangle_i^2$ 
can also be extracted from the fundamental matrix. 
One has \cite{snell}

\be
\label{fuvar}
\langle v\rangle_i=
\left((2\overline{\overline{\bf N}}-\overline{\overline{\bf I}})
\overline{{\bf T}}-\overline{\bf T}_{sq} \right)_i,\quad i=1,\ldots
N_s.
\ee
In this equation, the $N_s$-dimensional vector $\overline{{\bf T}}$
is given by

\be
\overline{{\bf T}}=\overline{\overline{\bf N}}\,\,\overline{\bf \Psi},
\ee 
where $\overline{\bf \Psi}$ is the 
$N_s$-dimensional vector with all entries equal to 1.
The vector $\overline{\bf T}_{sq}$ is obtained from 
$\overline{\bf T}$ by squaring all entries. The average 
$\langle v\rangle$ over the initial conditions 
is obtained by evaluating the weighted sum

\be
\label{wsumvar}
\langle v\rangle=\sum_{i=1}^{N_s} w_i \langle v\rangle_i.  
\ee

\section{1D lattice} 
\label{Sec2}

\subsection{Periodic boundary conditions}
\label{subper}

\begin{figure}[htbp]
\includegraphics[width=8cm,height=8cm]{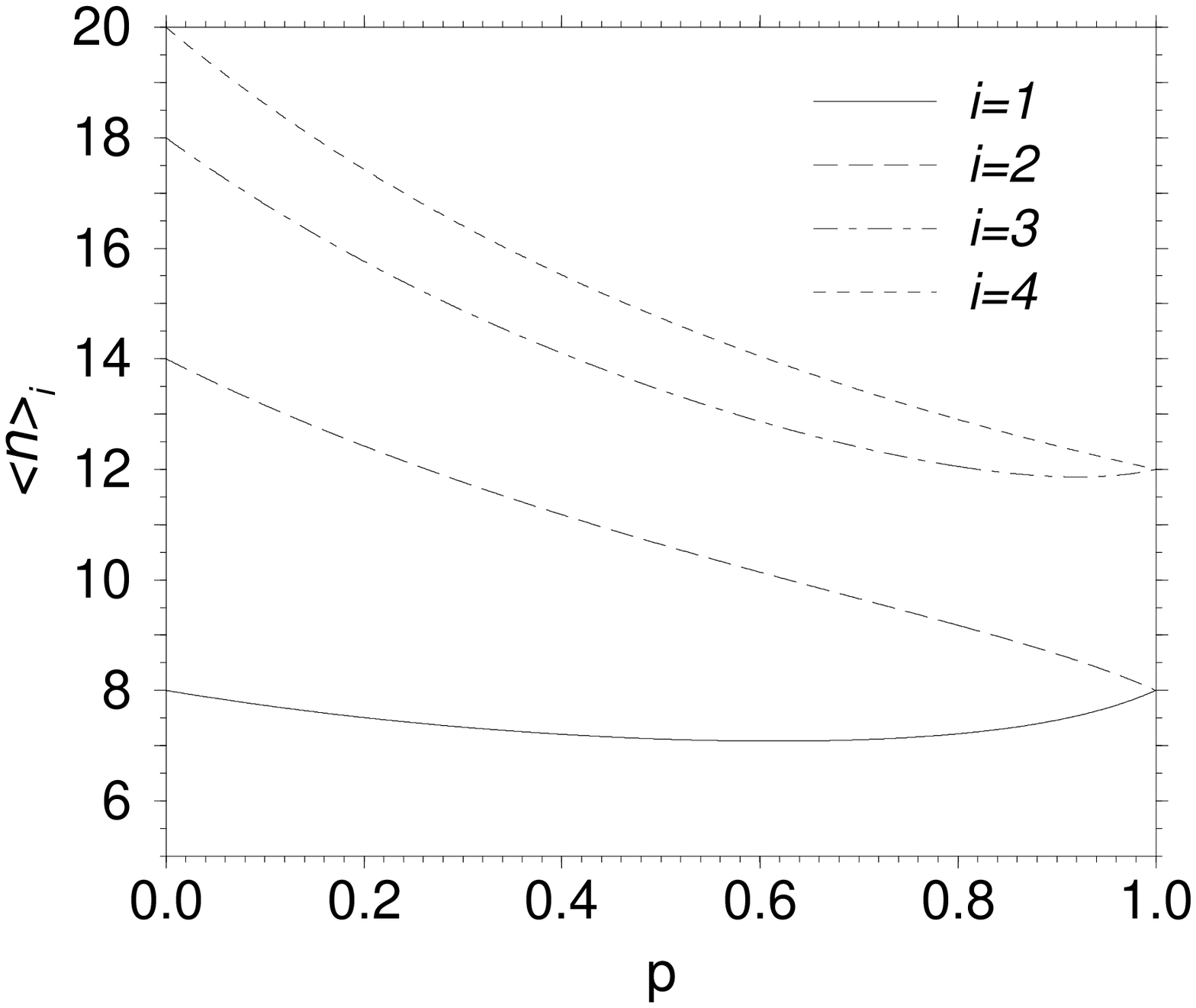}
\includegraphics[width=8cm,height=8cm]{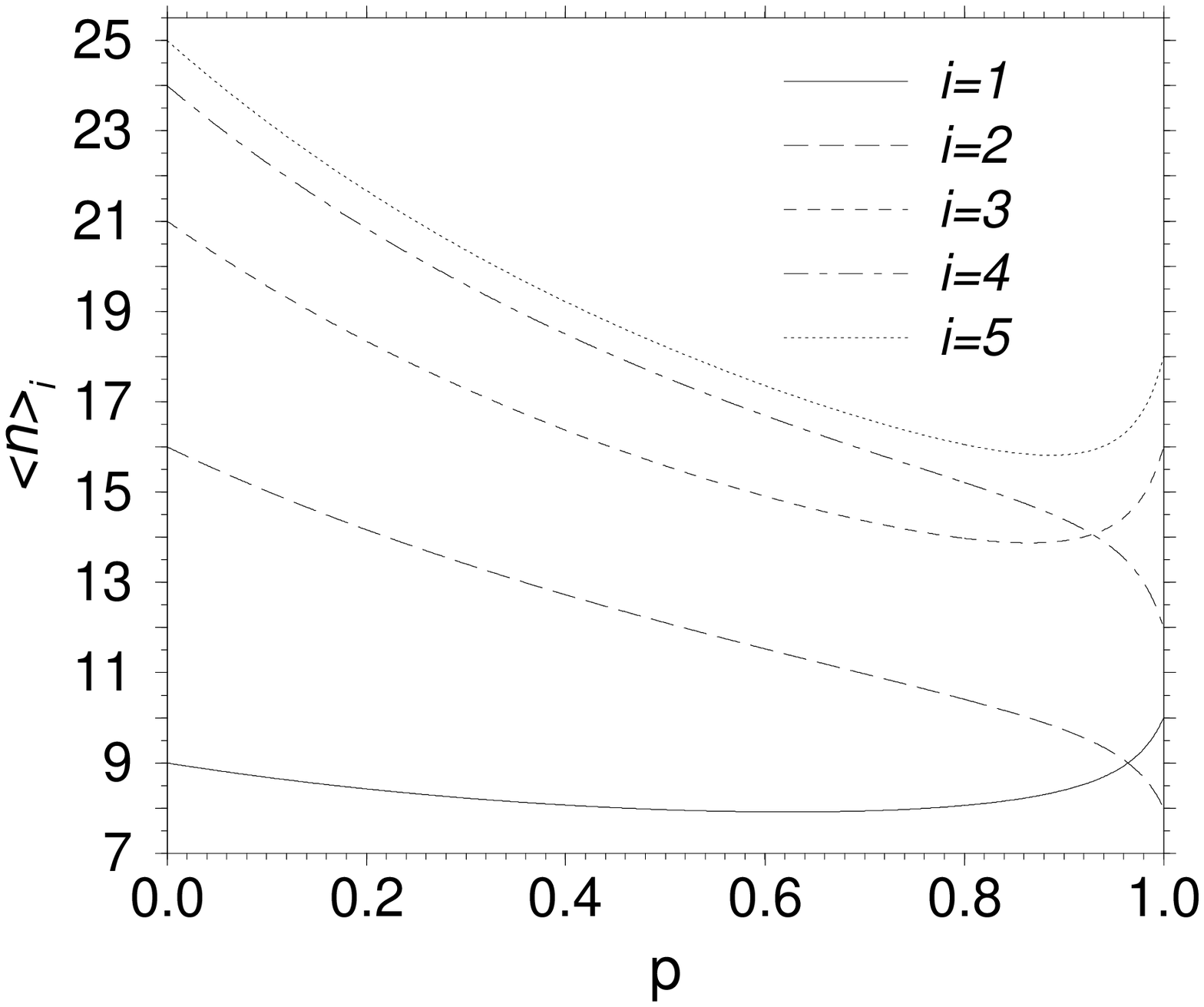}
\caption{\it \label{fig2} $p$-behavior of the local
reaction times in a $N=9$ case (left) and the $N=10$ case (right).}
\end{figure}

\begin{figure}[htbp]
\includegraphics[width=8cm,height=8cm]{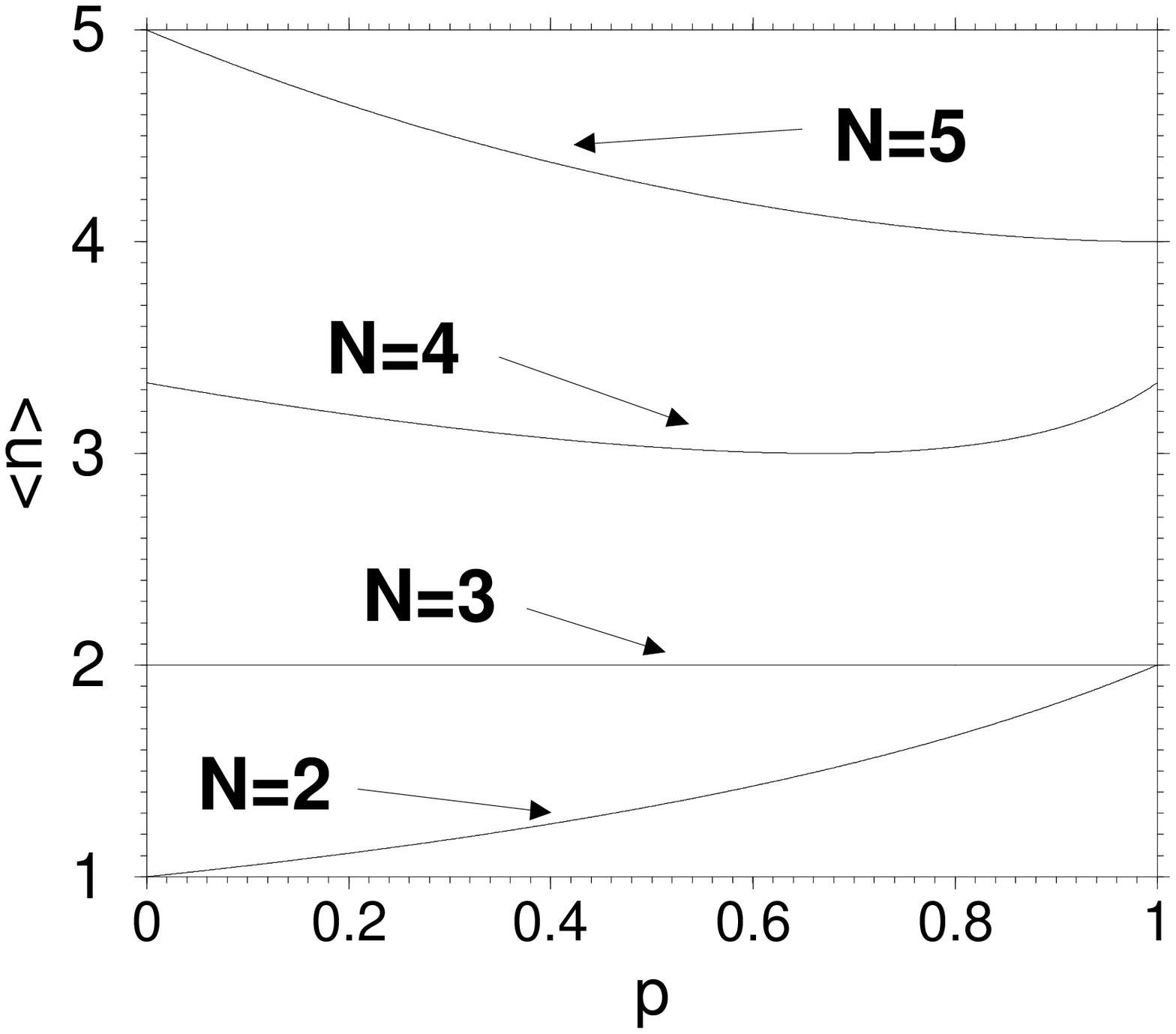}
\includegraphics[width=8cm,height=8cm]{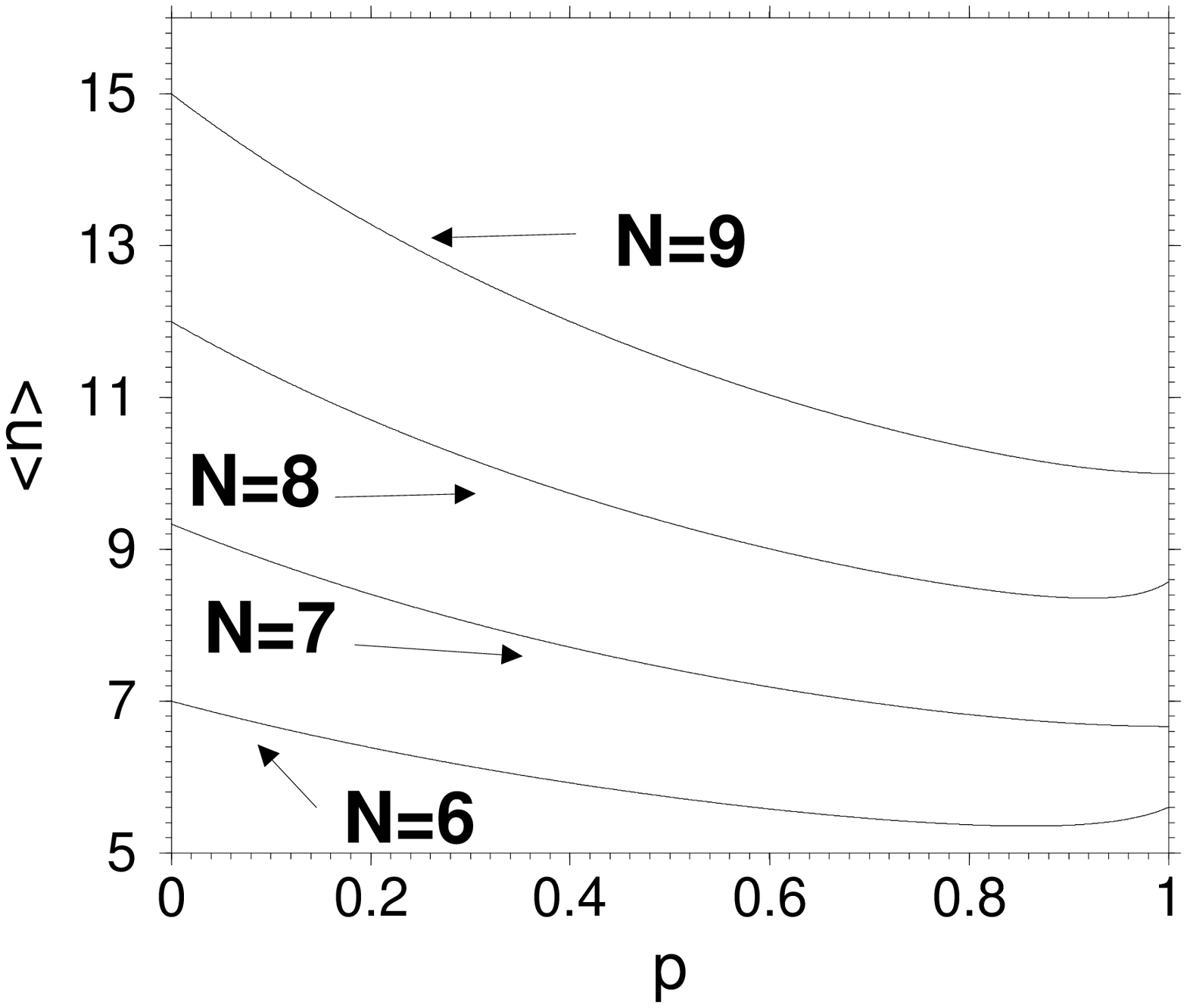}
\caption{\it \label{fig1} Global reaction time $\langle n\rangle$
as a function of 
$p$ for $N=2,\ldots,5$ (left) and $N=6,\ldots,9$ (right).}
\end{figure}

The Markov method can be used to recover results reported earlier 
\cite{eabad} using an approach based on backward difference equations. 
The first (rather remarkable) result
obtained at this level is that the local reaction times $\langle n\rangle_i$ 
(where the subscript $i$ is by construction 
equal to the initial interparticle distance
$d^0$) do not always decay monotonically with the mobility $p$, as one might
expect from an analysis of the relative motion
of the two walkers using a continuum approach. Indeed, 
monotonic decay of the $\langle n\rangle_i$'s 
only takes place for even values of 
$i$, whereas for odd values a local minimum is observed (see Fig. \ref{fig2}). 
This minimum shifts to higher $p$ values with increasing initial 
interparticle distance $d^0\equiv i$. 
The existence of such a minimum means that 
there is an anomalous inverted region in which  
$\frac{d\rt_i}{dp}>0$, implying that the efficiency of the reaction 
decreases with increasing temperature. 
Even though this behavior holds both for even and odd values of $N$ 
the minima are rather shallow in the odd lattice case 
and are therefore lost when one takes the average 
over the initial conditions to compute 
$\langle n\rangle$, in contrast to the even lattice case. 
Indeed, in computing $\rt$ as a function of $p$ via eq. 
$(\ref{elsum})$, one finds that  $\rt$
the behavior is different for even and odd lattices, 
,i.e., $\rt$ decays monotonically 
for odd values of $N$, while for even values a minimum 
of $\langle n\rangle$ for an intermediate value
$p=p_{min}^{\rt}$ is observed \footnote{This 
has been verified for lattice sizes up to N=36.}
(see Fig. \ref{fig1} and 
Table \ref{tabenct} for the detailed analytic expressions
of $\rt$ as a function of $p$). 
In other words, for $p>p_{min}^{\rt}$ one again observes 
an inverted region where $\frac{d\langle n\rangle}{dp}>0$. The value
of $p_{min}^{\rt}$ is strongly size-dependent and  
shifts to the right as $N$ becomes large, i.e.,
$\lim_{N\to\infty} p_{min}^{\rt}=1$ (see Table \ref{pmintab}). 

A qualitative explanation for 
the existence of a minimum in the case of an even lattice
has been given elsewhere \cite{procgran}. This argument suggests
that for lattices with an even number of sites the number of statistical
paths leading to prereactive configurations
with both particles at nearest neighbor sites is larger than in
the odd lattice case. In such a prereactive state, the next step will
lead to instantaneous reaction with a higher probability if it is 
asynchronous. A small degree of asynchronicity may thus diminish the
probability of mutual avoidance within the typical interaction zone 
and thereby result in an
increase of the reaction rate with respect to the fully synchronous 
case $p=1$. The
maximum efficiency will therefore correspond to a mixture of synchronous and
asynchronous steps. This 
effect is expected to wash out with increasing lattice size, since
``diffusional'' transport over long distances then becomes
the rate limiting step, and the latter is optimized by purely 
synchronous transport. Note also that a very similar argument may be
invoked to explain the shift of the minimum towards larger $p$-values 
observed in the curves $\rt_i(p)$ with increasing initial separation $i$.  
 
\begin{figure}[htbp]
\includegraphics[width=8cm,height=8cm]{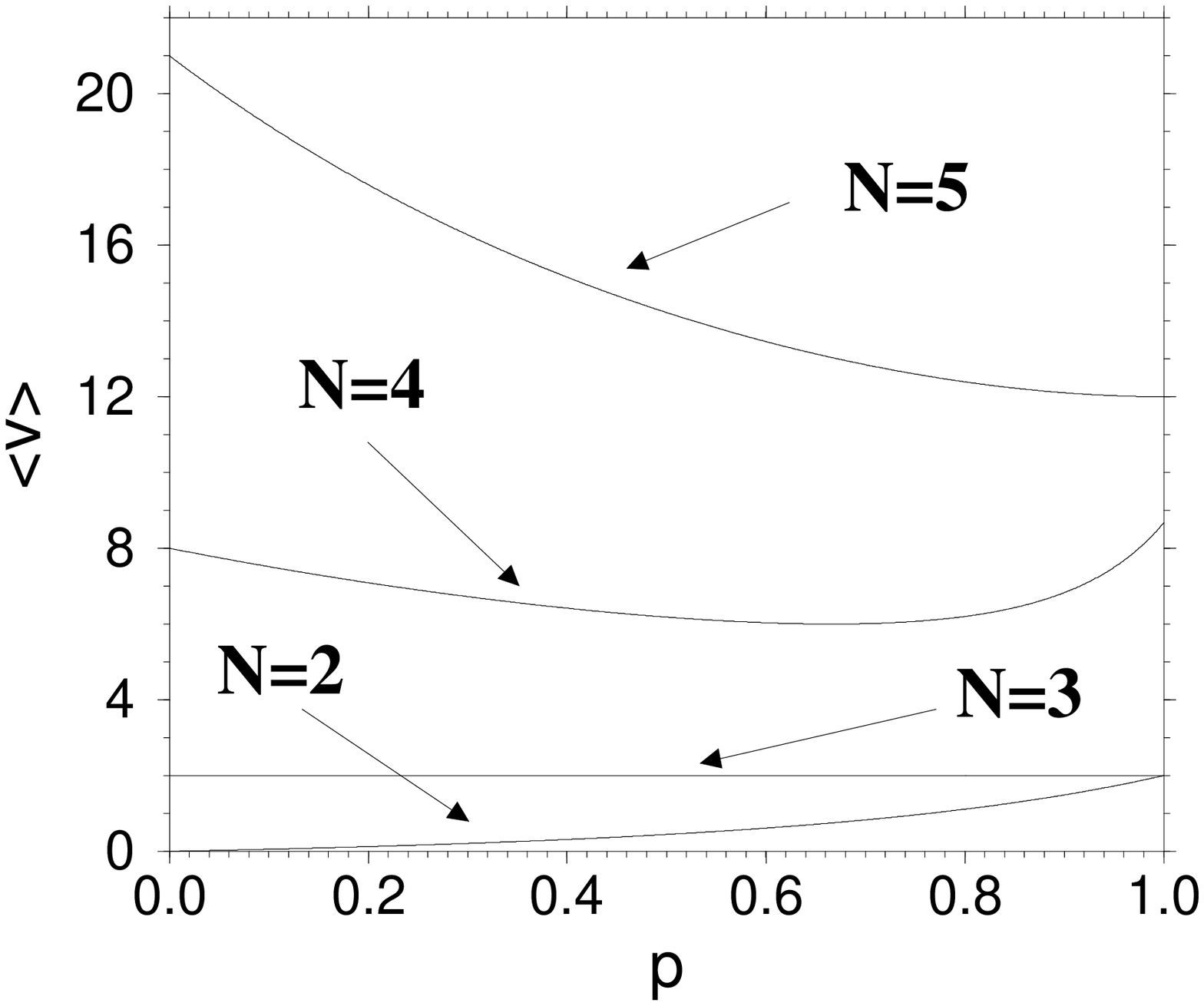}
\includegraphics[width=8cm,height=8cm]{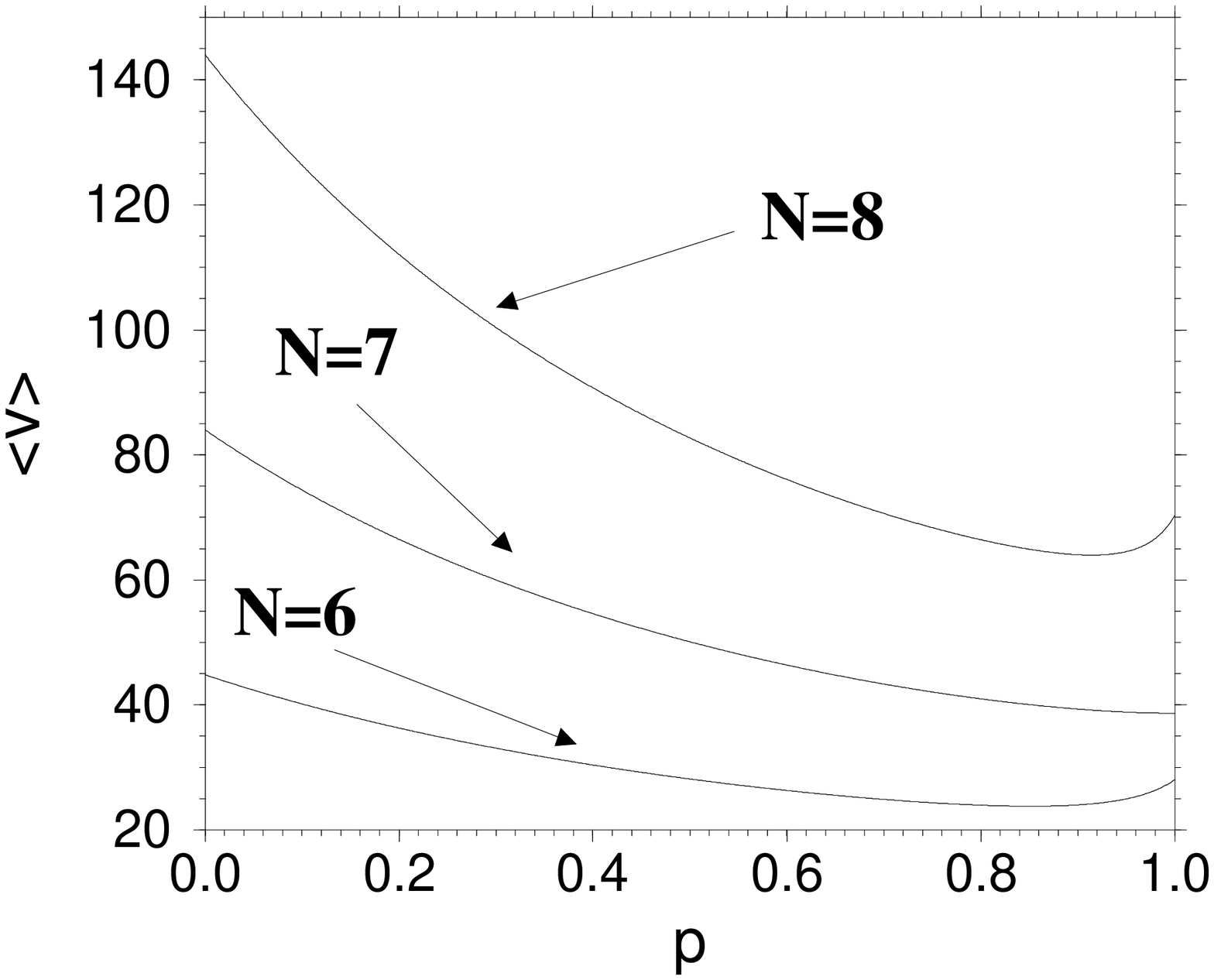}
\caption{\it \label{fig5} Global variance $\langle v\rangle$ 
as a function of $p$ for $N=2,\ldots,5$ (left) and $N=6,\ldots,8$ (right).}
\end{figure}

These results can be straightforwardly extended to higher-order
moments. In the 1D case, this is most naturally done 
via a generating function approach \cite{mont1, eabad2}, 
but the computation of the 
second-order moment is also possible in the framework of 
Markov theory (cf. eqs. (\ref{fuvar}) and (\ref{wsumvar})).
Remarkably, the variances $\vrt_i$ and $\vrt$ behave in 
a similar way as the first order 
moments. For example, the global variance $\vrt$ 
decreases monotonically with $p$ for odd lattices, 
whereas for even values of $N$ 
the curves $\vrt(p)$ display a minimum 
at a value $p=p_{min}^{\vrt}$ 
which again rapidly shifts towards one with increasing lattice
size (see Tables \ref{tabvar} and \ref{pmintab2}
and Fig. \ref{fig5}). One has $p_{min}^{\vrt}<p_{min}^{\rt}$
with the exception of the $N=4$ lattice, for
which both probabilities are equal (compare 
Tables \ref{pmintab} and
\ref{pmintab2}). Thus, the quantities 
$\langle v \rangle$ and
$\langle n\rangle$ cannot be minimized simultaneously
in general. 

In the limiting cases $p=0$ (where the heavy walker plays the role 
of a stationary trap) and $p=1$ (limit of identical walkers), it is
possible to recover previous results valid for arbitrary lattice
size $N$. In the framework of the Markov approach, these expressions 
for the reaction time and its variance can be constructed by identifying 
patterns in the (integer) values generated in calculating 
these quantities as a function of increasing $N$. In particular, one has  

\be 
\label{simpenct}
\langle n\rangle_i=i\,(N-i) 
\ee
in the $p=0$ case \cite{mont1} and 

\be
\label{compenct}
\langle n\rangle_i=-\frac{i^2}{2}+\frac{Ni}{2}+
\frac{1}{4}(1-(-1)^N)(-1)^{i}\,{i}+\frac{1}{4}
\left(N+\frac{(-1)^N+1}{2}\right)(1-(-1)^{i})
\ee
in the $p=1$ case, whereby $i$ can in both cases take any 
possible values of the initial interparticle distance $d^0$, 
i.e. $i =1,2,\ldots,[N/2]$. 

Note that when $N$ and $i$ take even values, one has 
$\langle n\rangle_i=i\,(N-i)/2$, implying that the reaction 
time takes half the value of the $p=0$ case, as one would expect 
from the result obtained in the continuum limit. 
If one now considers the average 
over a uniform ensemble of non-reactive configurations
one respectively has 

\be
\label{et1}
\langle n\rangle=\sum_{i=1}^{[N/2]}w_i \langle n\rangle_i= 
\frac{N(N+1)}{6} 
\ee 
for $p=0$ \cite{mont1,koz4} and 

\ba
\langle n\rangle &= N(N+1)(N+2)/(12(N-1))  &\qquad \mbox{$N$ even,}  
\nonumber\\
\label{et2}
&=(N+1)(N+3)/12  &\qquad \mbox{$N$ odd}
\ea
for $p=1$. According to Eqs. (\ref{et1}) and 
(\ref{et2}) the purely synchronous case
($p=1$) is more effective than the purely asynchronous case ($p=0$)
for all lattices with $N\ge 3$, as one would anticipate from the 
continuum approximation. 

Turning now to the variance, one has 

\be
\langle n^2\rangle_i-\langle n\rangle_i^2=
\frac{i^4}{3} -\frac{2N}{3}i^3+\frac{2}{3}i^2+
\frac{N(N^2-2)}{3}i-i^2(N-i)^2 
\ee

and

\begin{equation}
\label{onewvar}
\langle v\rangle=\frac{N(N+1)(N-2)(N+2)}{30}.
\end{equation}
when $p=0$. In the $p=1$ case the local variance is given by

\begin{eqnarray}
\langle n^2\rangle_i&=&\frac{1}{12}i^4-\frac{1}{6}Ni^3+
(\frac{1}{6}-\frac{1}{4}(N+1)(1-(-1)^i))i^2+(\frac{1}{12}N^3+\frac{1}{12}N+
\frac{1}{4}N^2)i +\frac{1}{8}+\frac{1}{3}N\nn \\
&&+
\frac{1}{24}N^3+\frac{1}{4}N^2-(\frac{1}{8}+\frac{1}{3}N+\frac{1}{24}N^3+
\frac{1}{4}N^2)(-1)^i-\frac{1}{4}(N+1)Ni(-1)^i
\end{eqnarray}  

for even values of $N$ and 

\begin{eqnarray}
\langle n^2 \rangle_i&=&\frac{1}{12}i^4-\frac{1}{6}(N+(-1)^i)i^3
+(\frac{1}{6}-\frac{1}{4}N(1-(-1)^i))i^2+(\frac{1}{12}N^3+
\frac{1}{4}N^2-\frac{1}{6}N)i\nn\\
&&\hspace{-.7cm} -\frac{1}{24}N+
\frac{1}{24}N^3+\frac{1}{8}N^2+(\frac{1}{24}N-\frac{1}{24}N^3-
\frac{1}{8}N^2)(-1)^i+(\frac{1}{4}N-\frac{1}{12})i(-1)^i
\end{eqnarray}

for odd values of $N$. Thus, the global variance becomes

\begin{equation}
\label{evenoddvar}
\langle v\rangle=
\left\{ \begin{array}{cc}
N(N+1)(N+2)(N^2+2N+2)/(120 (N-1)) & \qquad \mbox{for $N$ even, } \\
\label{twowvar}
(N+1)(N+3)(N^2+2N-5)/120 & \qquad \mbox{for $N$ odd. } 
\end{array} \right. 
\end{equation} 

\subsection{Confining boundary conditions}
\label{cobc}

 We now consider the behavior of a 1D system subject to confining
boundary conditions, and compare this with the behavior summarized 
above for periodic boundary conditions. The differences can already be
seen by examining the three simplest cases:  $N=4, N=5$ and $N=6$.

\begin{figure}[htbp]
\includegraphics[width=8cm,height=8cm]{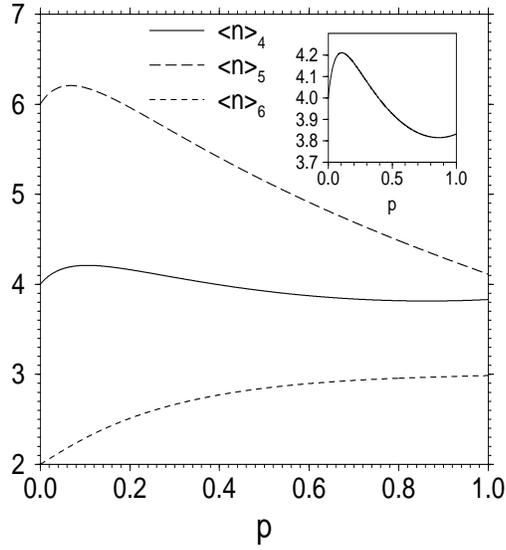}
\caption{\it \label{fig17} $p$-Behavior of the local reaction times
for those initial conditions 
characterized by the onset of an inverted region (1D lattice with $N=4$)
The inset clearly shows the double inverted region 
displayed by $\rt_4$.}
\end{figure}

In the $N=4$ case there are 6-symmetry distinct 
initial states in the $N=4$ case,
as already seen in \ref{defsym}. The initial states 1, 2 and 3
correspond to configurations with walker 2 at a boundary site.
In all three cases, the associated reaction times are found to
decay monotonically with $p$. In contrast, the $\rt_i(p)$ plots for 
the initial conditions 4, 5 and 6 display at least one inverted region 
where $\frac{d\rt_i}{dp}>0$ (see Fig. \ref{fig17}). 
The behavior for initial condition 4 is rather complex, i.e., 
the reaction time displays a maximum for small values 
of $p$ and a minimum for large
values of $p$, thereby giving rise to a double inverted region,
as shown in the inset of Fig. \ref{fig17}.
On the other hand, $\rt_5$ rapidly reaches a maximum for small values of $p$
and then decays sharply with $p$. Finally, $\rt_6$ 
increases monotonically for all $p$, i.e.,
 the inverted region comprises the whole
$p$-interval. We thus see that the behavior of the efficiency in the confining
case is far more complex than in the periodic case, where only monotonically
decreasing curves or curves with a single minimum are observed.

\begin{figure}[htbp]
\includegraphics[width=8cm,height=8cm]{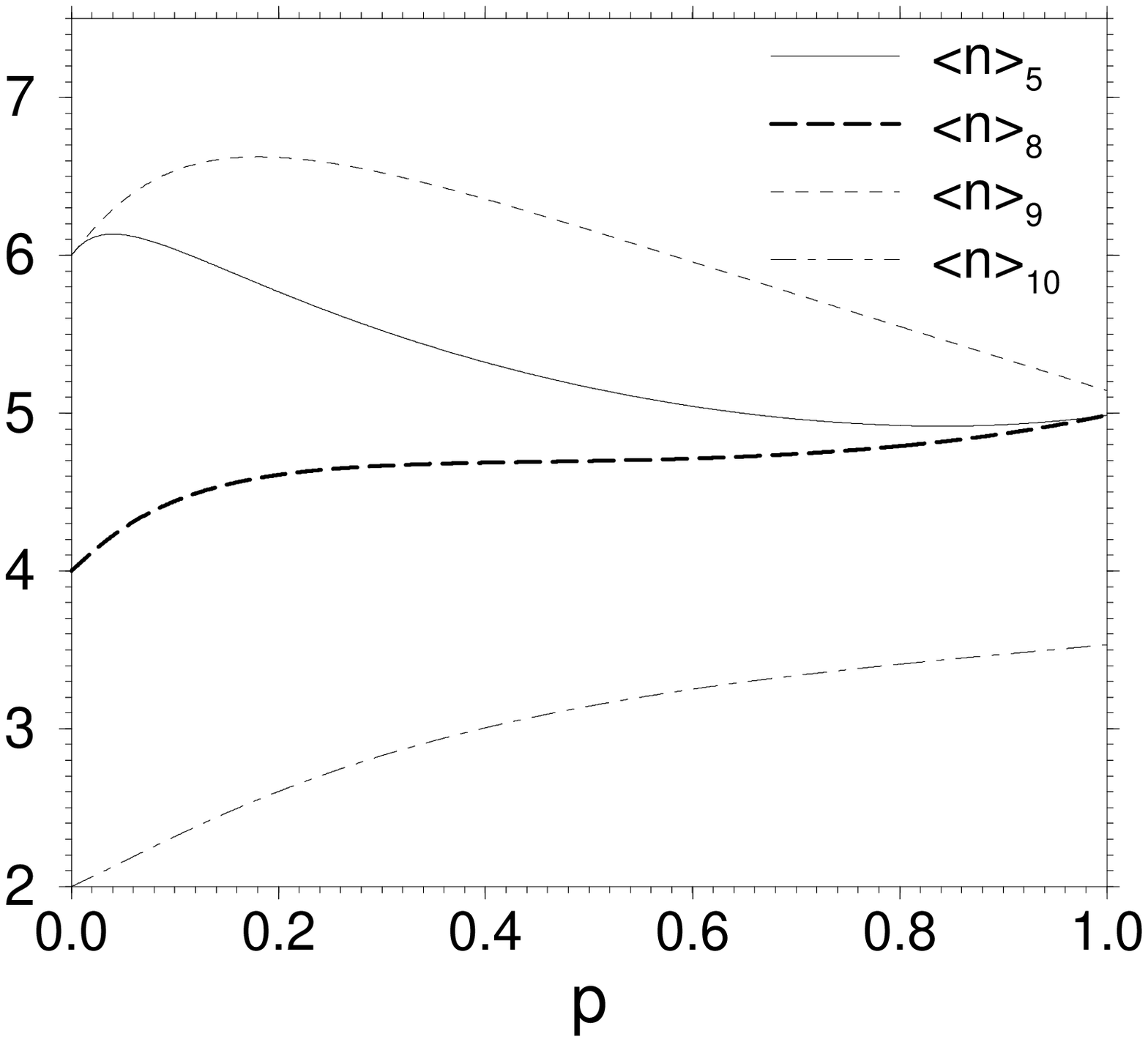}
\includegraphics[width=8cm,height=8cm]{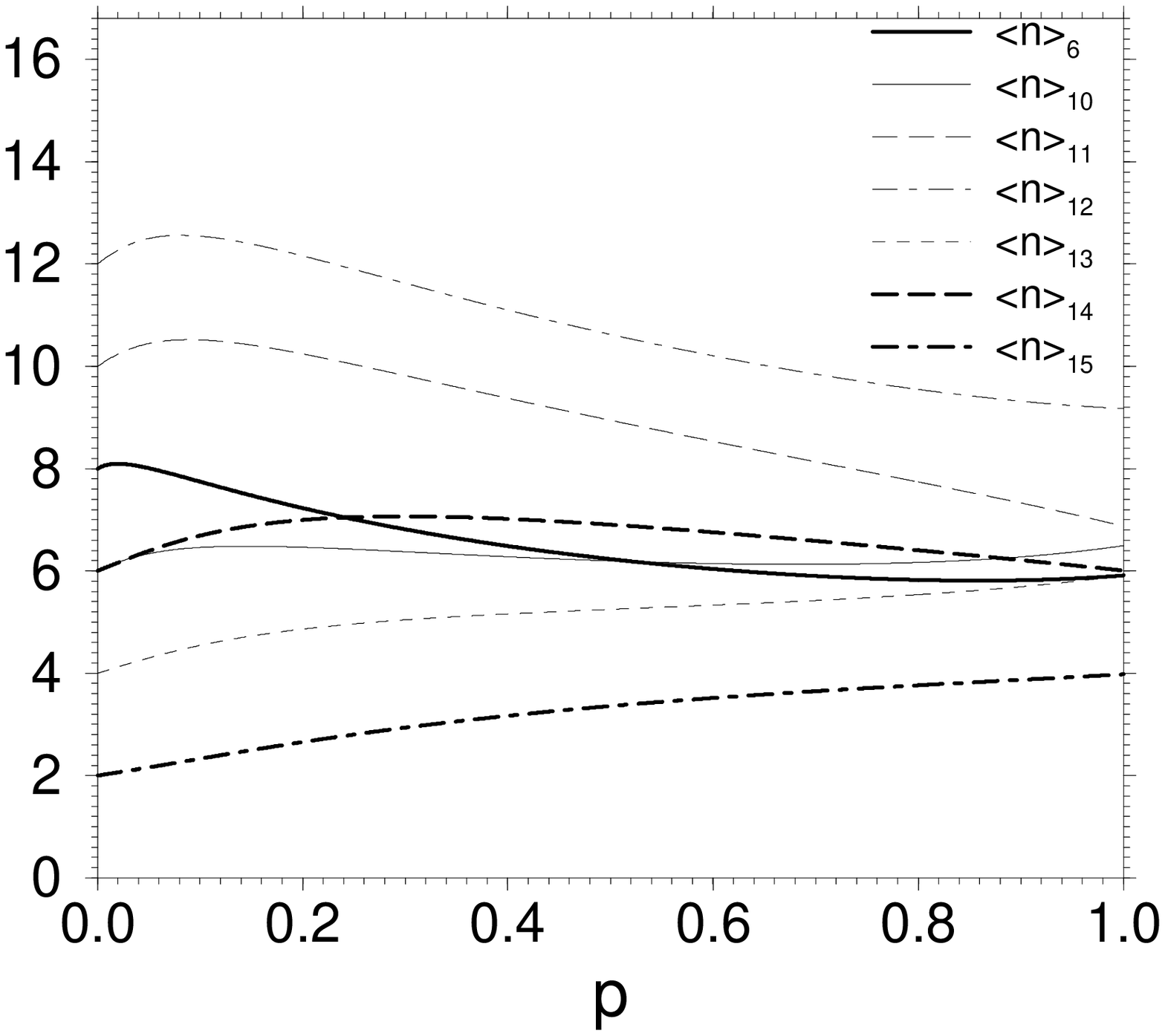}
\caption{\it \label{fig18} $p$-Behavior of the local reaction times
for some of the initial conditions 
characterized by the onset of an inverted region. The left (right)
graph corresponds to a 1D lattice with $N=5 (N=6)$.}
\end{figure}

In the cases $N=5,6$ a similar qualitative behavior is observed. 
More specifically, the initial states $1$ to $N-1$  
correspond to monotonically decreasing reaction times, 
while the initial states $N$ to $N_s$
are associated with $\rt_i(p)$ curves displaying inverted regions
with positive $p$-derivative. In some
cases the inverted regions extend over a very short $p$ interval, 
this is e.g. the case 
for the initial states 6 and 7 in the $N=5$ lattice and the initial 
states 7, 8 and 9 in the
$N=6$ case, whose associated $\rt_i$'s 
display a single maximum very close to $p=0$.
The other reaction times displaying anomalous behavior 
are plotted in Fig. \ref{fig18}. It is noteworthy that
those initial conditions for which a double inverted region is observed
correspond to contiguous walker positions, whereby  
the heavy walker is placed at an interior site 
closer to the boundary or at most 
at the same distance from the boundary as the light walker. 
In contrast, those configurations with contiguous 
particles in which the light walker is closer to the boundary are 
associated with monotonically increasing $\rt_i(p)$ profiles,
 as one might intuitively expect (for large $p$ 
the heavy walker can move away from the light walker more easily 
leaving a larger space between the boundary and itself available for the 
motion of the light walker). Finally, all 
the other initial conditions giving rise to anomalous behavior 
correspond to $\rt_i(p)$ profiles displaying a single maximum 
as a function of $p$.

\begin{figure}[htbp]
\includegraphics[width=8cm,height=8cm]{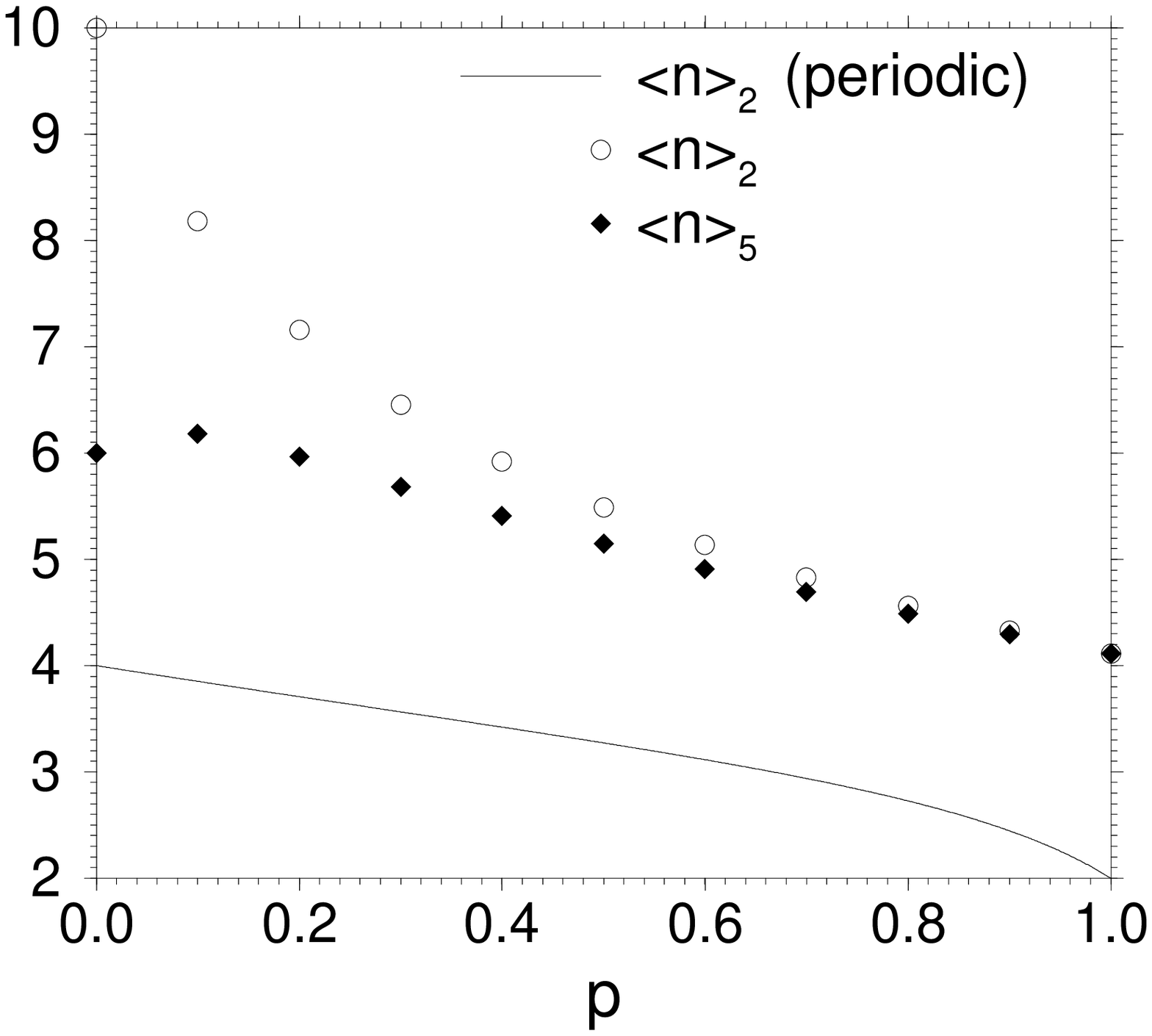}
\includegraphics[width=8cm,height=8cm]{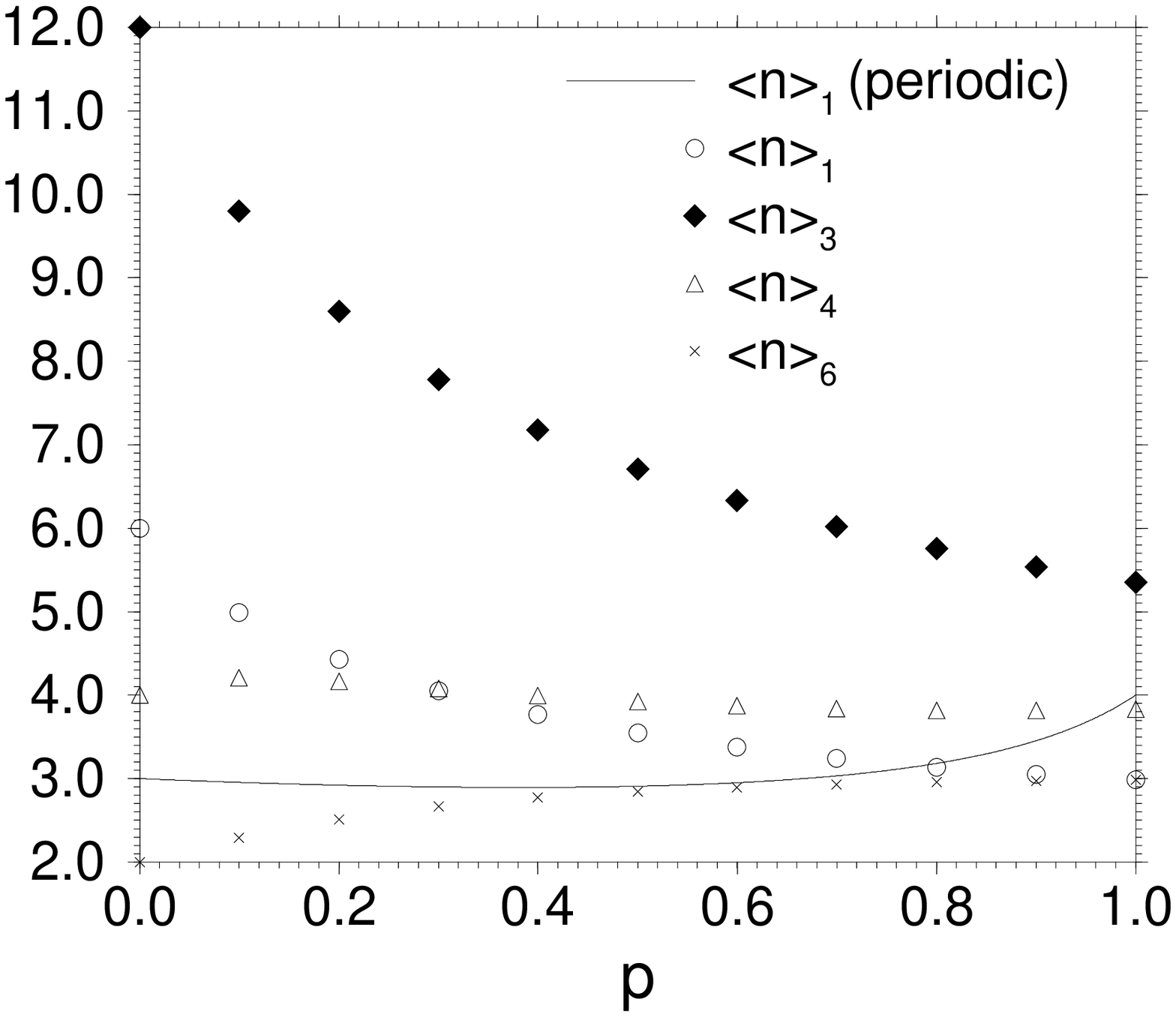}
\caption{\it \label{confvsper} $p$-behavior of the local reaction
times in the confining and the periodic $N=4$ case. 
The initial states 2 and 5 in
the confining lattice collapse to the initial state 2 
(corresponding to walkers at next to nearest sites 
in the periodic lattice) when the lattice boundary sites 
are connected to each other (see left graph).
The initial states 1,3, 4 and 
6 in the confining lattice collapse to the initial state 1 
(corresponding to contiguous walkers
in a periodic lattice) when the lattice boundary sites 
are connected to each other (see right graph).}
\end{figure}

Let us now compare the above behavior of the local reaction time with
our findings for the $N=4$ periodic case. To begin with, note that
if one switches from confining to periodic
boundaries, States 2 and 5 become equivalent to, 
say, State 2 in the periodic case. Fig. \ref{confvsper} (left) displays
a comparative plot for the $\rt_i(p)$ curves in these cases. 
It is seen that for all values of $p$
the reaction times $\rt_2$ and $\rt_5$ are well below the 
corresponding values of $\rt_2$ for a periodic lattice. 
On the other hand, states 1, 3, 4 and 6 become equivalent to 
state 1 when switching to periodic boundary conditions. 
As seen in Fig. \ref{confvsper} (right)  
the reaction times  $\rt_{1,3,4}$ in the confining
system are larger than the corresponding reaction time
$\rt_1$ in the confining case for most values of $p$. Exceptions are
found for $\rt_1$ and $\rt_4$, which become smaller 
in the high-$p$ regime. In contrast, the initial state 6 is the only 
one in which for any ``temperature'' the reaction is more efficient 
in the confining than in the periodic case.  

In the cases $N=5,6$, one has a similar behavior of $\rt_1$ and
$\rt_{N_s}$, i.e., there is a significant $p$ interval where $\rt_1$
becomes smaller than in the periodic case, whereas this interval 
extends to the full physical range [0,1] if one starts from the
$N_s$-th initial configuration, in which the light walker is placed at a
boundary site next to the heavy walker. 

\begin{figure}[htbp]
\includegraphics[width=8cm,height=8cm]{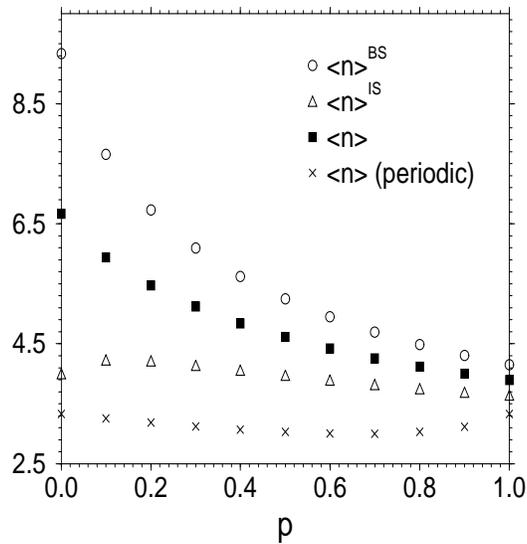}
\caption{\it \label{globeh} $p$-Behavior of the global
reaction times for the periodic and the confining $N=4$ case.}
\end{figure}

Finally, we turn to the difference in behavior between the
confining and the periodic case at the level of the global 
reaction time. In the confining case, 
the weights $w_i$ used to compute $\rt$ take a very 
simple form, i.e., $w_i=N_s^{-1}=2/N(N-1)$. One thus has 
$\rt=\frac{2}{N(N-1)}\sum_{i=1}^{N(N-1)/2} \rt_i$. In addition, 
one can also define other global averages by averaging over subsets
of initial conditions, e.g. 

\be 
\label{subglav}
\rt^{x_2^0=i}=\frac{1}{N-1}\sum_{x_1^0=1 x_1^0 \neq i}^N \rt_{(x_1^0,i)}
\ee 
is the average over a homogeneous subset of initial conditions with
a fixed value of the heavy walker's initial position.
(In the periodic case these averages are clearly
$i$-independent and equal to the all averaged reaction time $\rt$). 
In the case of a $N=4$ lattice, there are only two 
such symmetry-distinct averages, namely $\rt^{x_2^0=1}\equiv
\rt^{BS}$ and $\rt^{x_2^0=2}\equiv \rt^{IS}$ with the heavy
walker respectively placed at a boundary site (BS) and an interior site
(IS) initially. For an arbitrary lattice size $N$, these global 
reaction times can be straightforwardly calculated from the
local reaction times for each symmetry-distinct initial condition 
by a suitable weighting of the individual contributions. In the
(particularly simple) case of the $N=4$ lattice, all weights are 
equal, i.e.,

\ba
\rt^{BS}=\frac{\rt_1+\rt_2+\rt_3}{3}, \\
\rt^{IS}=\frac{\rt_4+\rt_5+\rt_6}{3} 
\ea
As it turns out $\rt^{BS}$ decreases 
monotonically with $p$, while $\rt^{IS}$ first increases and then
decreases (see Fig. \ref{globeh}). Notice that the inequality 
$\rt^{BS}>\rt^{IS}$ holds for all values of $p$. 
In addition, one finds that $\rt^{BS}$,$\rt^{IS}$
and thus also $\rt=\frac{\rt^{BS}+\rt^{IS}}{2}$ are larger than the
global reaction time $\rt$ for
the periodic case, thereby emphasizing that at this level of 
description the confining system
is less efficient over the whole temperature range. In the cases 
$N=5,6$ a similar behavior of the global reaction
times is found, i.e., for a given $p$ the global reaction times 
decrease when the heavy walker is shifted to the interior of the 
lattice, but in all cases they remain larger than in the 
periodic case.    

As for the $p$-dependence of the global reaction times
in the $N=5,6$ cases, the reaction time $\rt^{BS}\equiv \frac{1}{N-1}
\sum_{i=1}^{N-1} \rt_i$ is found to decrease monotonically with $p$, 
whereas the other global reaction times (corresponding to configurations
where the heavy walker starts from an interior site) first increase 
with $p$, thereupon reach a maximum for $p$ well below $0.5$ and finally 
decrease 
for larger $p$ values. These maxima are again 
lost in the ``all averaged'' reaction time $\rt$, 
which decreases monotonically with $p$ for all lattices in the
confining case, in contrast with the even-odd effect reported 
for periodic lattices. 

\section{2D lattice}
\label{Sec4}

\subsection{Periodic boundary conditions}
\label{2dp}

As in the 1D case, one can take advantage of symmetry to simplify 
the formulation of the problem, here by lumping the $N(N-1)$
possible non-absorbing initial states $(x_1^0,x_2^0)$ in a square planar
lattice into a set of $N_s$ states.   
Each symmetry-distinct initial state is now characterized by the 
horizontal distance $d_x^0$ 
and the vertical distance $d_y^0$ between the walkers (measured in lattice
spacings) along the two orthogonal 
directions $x$ and $y$ specified by the vector basis of the
elementary lattice cell. Both $d_x^0$ and $d_y^0$ may take values from $0$ to 
$[\sqrt{N}/2]$, but the absorbing state with both walkers at the same site
($d_x^0=d_y^0=0$) is excluded. Taking advantage of exchange symmetry with 
respect to $d_x^0$ and 
$d_y^0$, one can therefore also exclude configurations for which, 
say, $d_y^0>d_x^0$. Thus, the total number of symmetry-distinct states 
is \footnote{In the following equation $CR_A^{\, B}$ stands for the 
number of 
combinations with repetition of $A$ elements choose $B$.}   
$N_s=CR_{[\sqrt{N}/2]+1}^{\,\, 2}-1={[\sqrt{N}/2]+2 \choose 2} -1$. To remain
consistent with the numbering criterion used in the 1D case, we shall
number these states from 1 to $N_s$ by increasing Euclidean distance
$d^0=\sqrt{(d_x^0)^2+(d_y^0)^2}$. For example, for a $3\times 3$ lattice
one has 2 symmetry-distinct states, namely:

\[
\begin{array}{ccccccccc}
\mbox{State 1} &\qquad &\mbox{State 2}\\
&&\\
2-1-0 &\qquad & 2-0-0\\
|\hspace{0.6cm} |\hspace{0.6cm} | &\quad & |\hspace{0.6cm} 
|\hspace{0.6cm} | \\
0-0-0 & \qquad & 0-1-0 \\
|\hspace{0.6cm}  | \hspace{0.6cm} | &\qquad &  | \hspace{0.6cm} | 
\hspace{0.6cm}| \\
0-0-0 &\qquad & 0-0-0
\end{array}
\]

In order to compute the reaction time and its variance, one now proceeds 
as in the 1D case, i.e., one first computes the corresponding 
$N_s\times N_s$ fundamental matrix and then respectively applies the 
equations (\ref{locrt}) and (\ref{fuvar}) for the local quantities
$\rt_i$ and $\vrt_i$ and eqs. (\ref{elsum}) and 
(\ref{wsumvar}) for the global averages $\rt$ and $\vrt$, whereby 
the weights $w_i$ are again straightforward to compute.   

\begin{figure}[htbp]
\includegraphics[width=8cm,height=8cm]{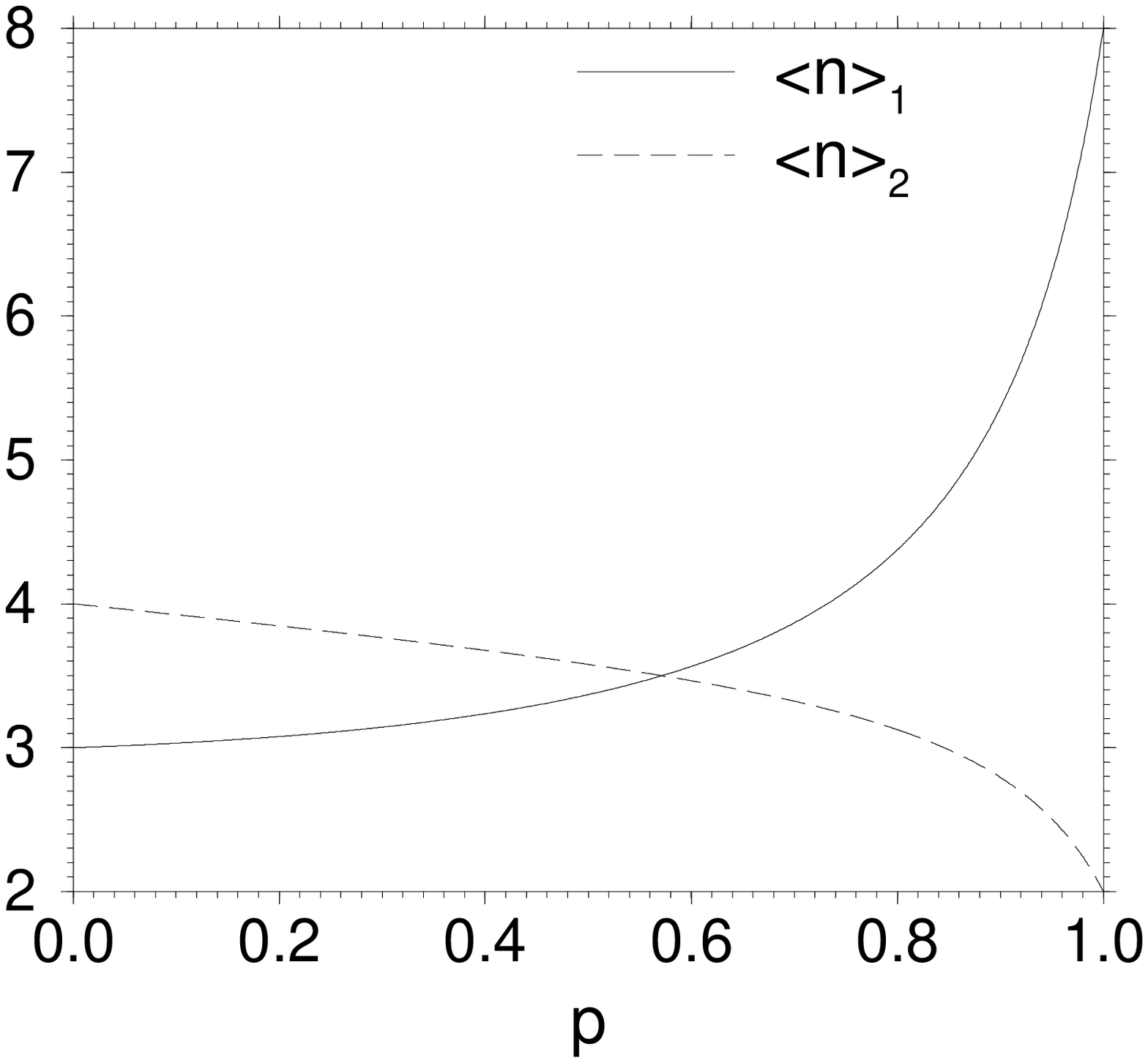}
\includegraphics[width=8cm,height=8cm]{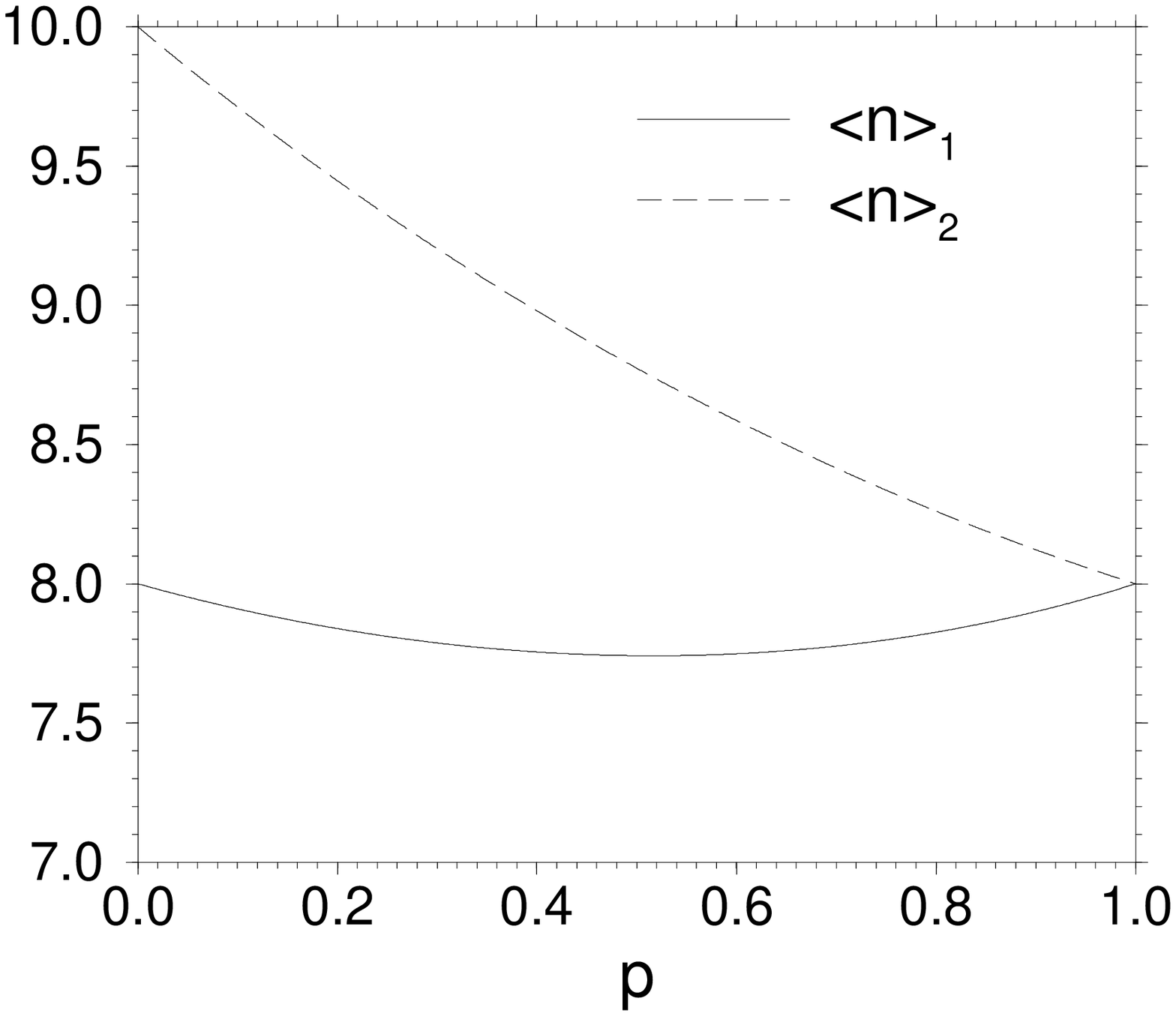}
\caption{\it \label{fig8} $p$-Behavior of the local reaction times
$\rt_1$ and $\rt_2$ in a $2\times 2$ (left) and a
$3\times 3$ periodic square planar lattice (right). 
The values of the effective distance $\hat{d^0}$ associated with the
initial state 1 and 2 are respectively 1 and 2.}
\end{figure}

Let us first discuss the behavior of 
the individual contributions $\langle n\rangle_i$ 
to the global reaction time $\langle n\rangle$.
In the case of a $2\times 2$ lattice, 
there are only two possible symmetry-distinct
initial states, i.e., State 1 is a configuration in which the walkers
are placed at nearest neighbor sites, whereas in State 2 is 
they are placed diagonal to each other. 
Fig. \ref{fig8} (left) displays the behavior of the 
reaction times $\langle n\rangle_1$ and $\langle n\rangle_2$   
as a function of $p$. While  $\langle n\rangle_1$ decreases with 
increasing $p$, $\langle n\rangle_2$
increases with $p$ at a slow rate for low values of $p$ and more
rapidly for larger values. In the case of a $3\times 3$ lattice, 
the behavior of the reaction time is
also different for each of the two symmetry-distinct initial conditions
$1$ and $2$, i.e., $\langle n\rangle_1$ displays a shallow
minimum as a function of $p$, while $\langle n\rangle_2$ is a
monotonically decreasing function (see Fig. \ref{fig8}, right). 

\begin{figure}[htbp]
\includegraphics[width=8cm,height=8cm]{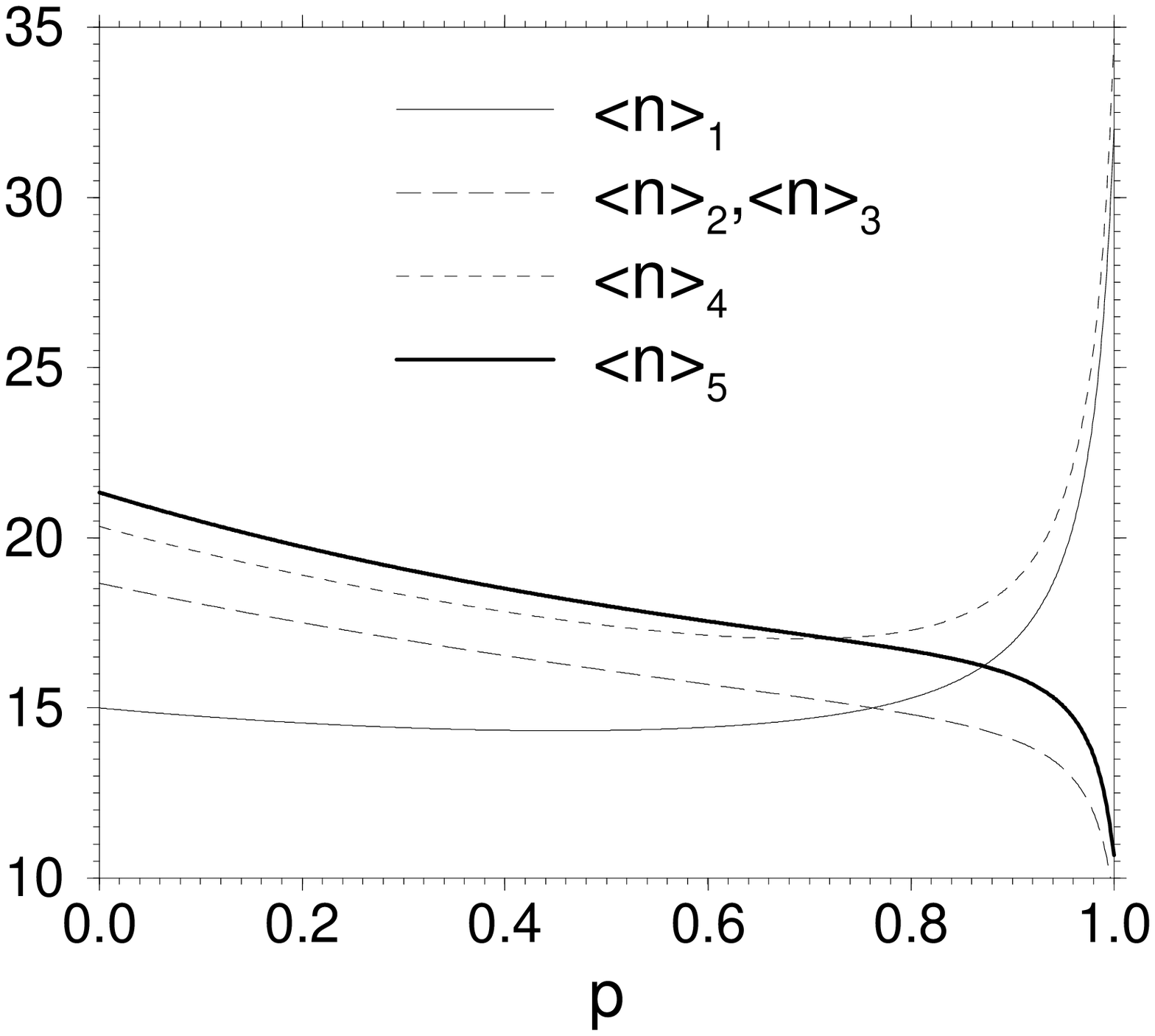}
\includegraphics[width=8cm,height=8cm]{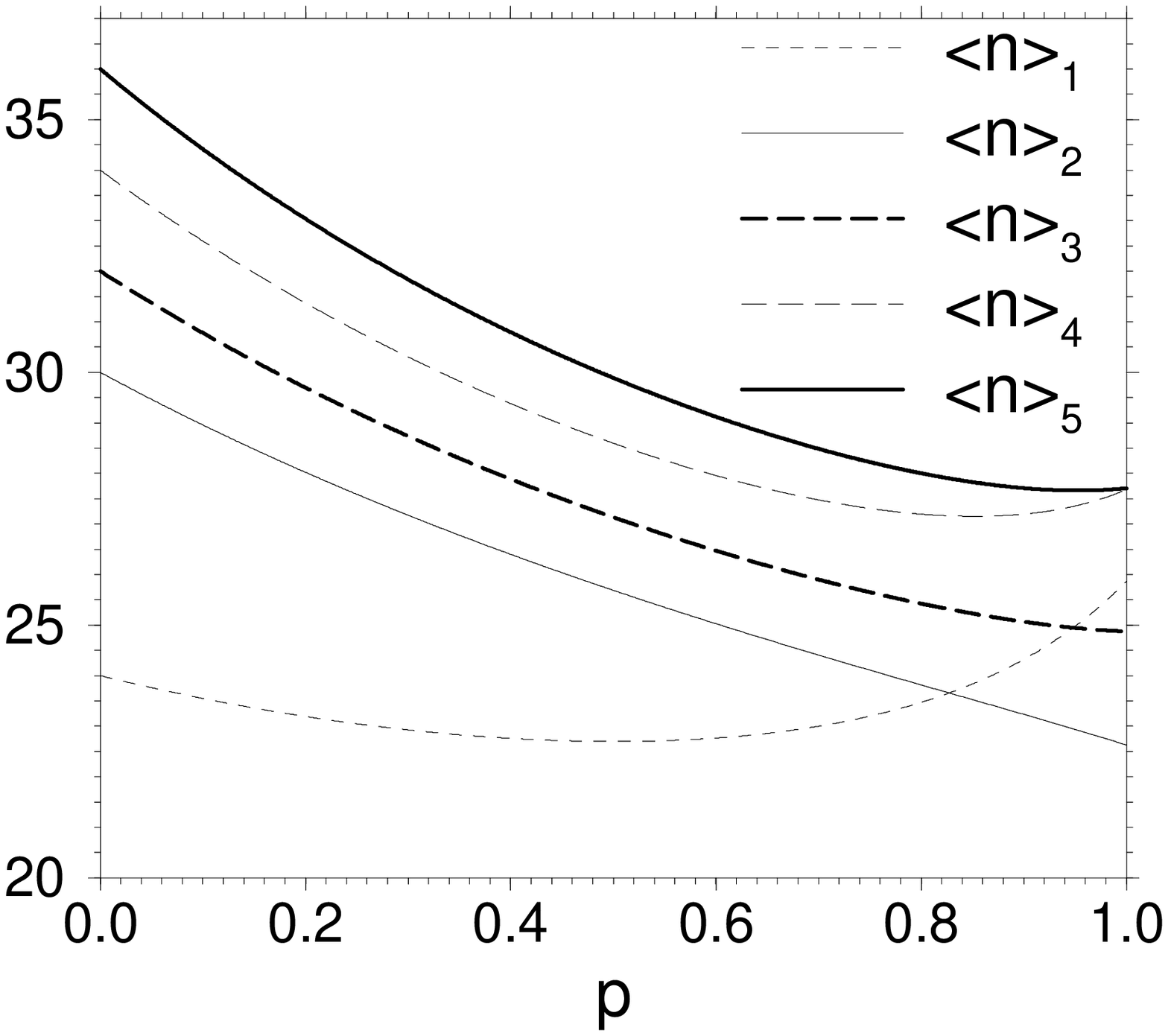}
\caption{\it \label{fig10} $p$-Behavior of the local reaction times 
for each of the five symmetry-distinct initial configurations
in a $4\times 4$ square planar lattice (left) and a $5\times 5$ 
square planar lattice (right). In both cases the values of
$\hat{d^0}$ associated with the initial states 1, 2, 3, 4 and 5 are
respectively 1, 2, 2, 3 and 4.}
\end{figure}

\begin{figure}[htbp]
\includegraphics[width=10cm,height=10cm]{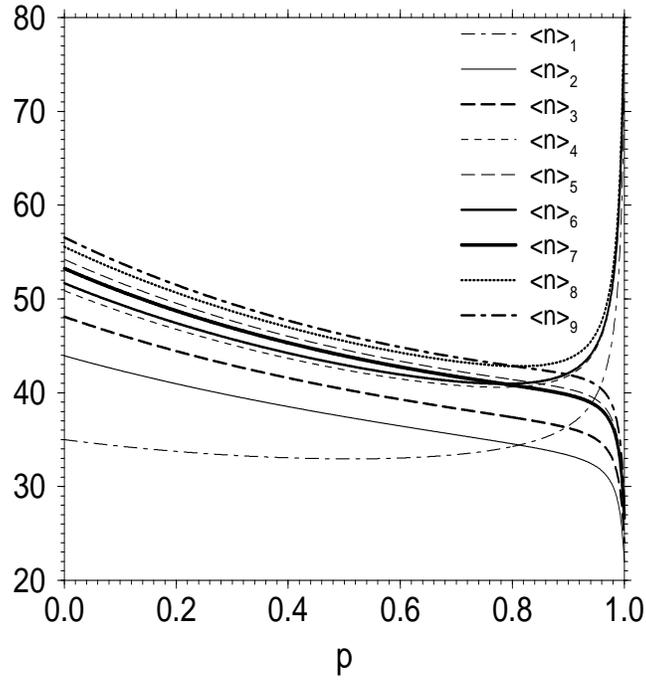}
\caption{\it \label{fig12} $p$-Behavior of the local reaction times 
for each the 9 symmetry-distinct initial configurations
in a $6\times 6$ square planar lattice. The values of 
$\hat{d^0}$ associated with the initial states 1, 2, 3, 4, 5, 6, 7, 8 and 9
are respectively 1, 2, 2, 3, 4, 3, 4, 5 and 6.}
\end{figure}

A careful examination of the behavior for somewhat larger periodic 
square planar lattices ($ 16 \le N \le 36$) suggests the following rule
of thumb: those initial states for which the effective distance 
$\hat{d^0}\equiv d_x^0+d_y^0$ is even 
correspond to monotonically decreasing $\rt_i(p)$ 
profiles, while the
profiles $\langle n\rangle_i(p)$ associated with 
odd values of $\hat{d^0}$ show an inverted region. Each of these 
inverted regions is associated with a local minimum with respect to $p$, 
except in the $2\times 2$ lattice case, where the inverted region 
extends over the whole $p$ interval.  
As expected, the minimum shifts to the right with increasing 
value of $\hat{d^0}$ 
(see Figs. \ref{fig10} and \ref{fig12} for the cases of a 
$4\times 4$, $5\times5$ and $6\times 6$ torus). This even-odd behavior 
in terms of $\hat{d^0}$ is analogous to the even-odd behavior 
in terms of $d^0$ observed in 1D (cf. Fig. \ref{fig2}). 

Departures from the above rule become more frequent with
increasing lattice size. An example is given
by the initial state 5 corresponding to the $5\times 5$ square planar 
lattice, which
displays a tiny region with $\frac{d\rt_5}{dp}>0$ and
a very shallow minimum near $p=1$, as a careful inspection
of the thick solid line in the right graph of Fig. \ref{fig10} 
would reveal. Shortly we shall see that as a result of this 
inverted regions at the level of $\rt$ also arise
for sufficiently large odd values of $N$, in contrast with the 1D case. 

\begin{figure}[htbp]
\includegraphics[width=8cm,height=8cm]{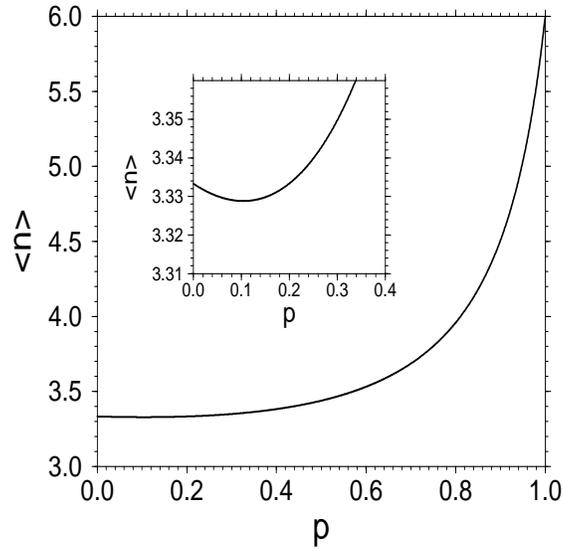}
\caption{\it \label{fig7} $p$-behavior of $\langle n\rangle$ 
in an $N=4$ periodic square planar lattice. The inset
shows the minimum of the curve for $p\approx 0.104$.}
\end{figure}

\begin{figure}[htbp]
\includegraphics[width=8cm,height=8cm]{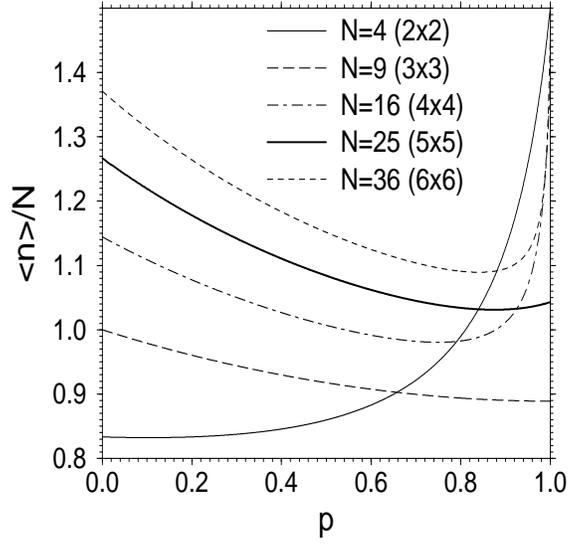}
\caption{\label{globeh2} $p$-Behavior of the global reaction time (divided
by the lattice size) for 2D square planar lattices of increasing size.}
\end{figure}

Next, let us analyze the $p$-behavior of $\rt$ given in Table \ref{n2d}. 
To this end, we first focusing on the case of the smallest possible lattice 
($2\times 2$ lattice, corresponding to $N=4$).  
Fig. \ref{fig7} displays the behavior of the global reaction
time $\langle n\rangle$ as a function of $p$ 
for a $2\times 2$ periodic square lattice. The inset shows that 
there is a small region of $p$-values where 
$\frac{d\langle n\rangle}{dp}<0$. However, if one compares this 
plot to the $N=4$ plot depicted in Fig. \ref{fig5} (left), 
it is seen that the inverted region with 
$\frac{d\langle n\rangle}{dp}>0$ is much broader in the 
case of a $2\times 2$ lattice than in the case of a $N=4$ ring, 
leading to a smaller value of $p_{min}^{\rt}$ in the $2D$
case. This behavior persists if one considers 
larger lattices with an even number of sites
(compare e.g. the values of $p_{min}^{\rt}$ for $N=16$ and $N=36$ 
in Tables \ref{pmintab} and \ref{pmintab2d}). 
We thus conclude that for fixed $N$ the inverted region 
$[p_{min}^{\rt},1]$ observed for even lattices in 1D is 
considerably extended in 2D.  
Moreover, in contrast to the 1D case an inverted region 
also appears in the $5\times 5$ lattice case
(see the corresponding plot in Fig. \ref{globeh2}), 
and numerical evidence suggests that inverted regions also arise  
in the $7\times 7$ and the $9\times 9$ cases \cite{eabad}.  

There is another respect in which the anomalous $p$-behavior of $\rt$
is here enhanced with respect to the 1D case.
Let us consider the relative enhancement of the efficiency 
$\Delta$ in the case of an even lattice, defined as follows: 

\be 
\Delta\equiv
\frac{1}{\rt(p=p_{min}^{\rt})}\int_{p_{min}^{\rt}}^1 \frac{d\rt}{dp}\,dp
=\frac{\rt(p=1)-\rt(p=p_{min}^{\rt})}{\rt(p=p_{min}^{\rt})}
\ee
From Table \ref{compdelta}
it is seen that this quantity is very large with respect to the 
1D case, even for large
lattices ($N\ge 16$). On the other
hand, the interval $[p_{min}^{\rt},1]$ corresponding to 
the inverted region contracts dramatically with increasing lattice 
size, as seen from Table \ref{pmintab2d} and expected from
the arguments for the 2D case. Thus, a minute 
amount of asynchronicity $\delta p<0$ brings about a drastic 
decrease in $\rt$ with respect to the fully synchronous 
case $p=1$.    

We have also studied the global variance $\vrt$ in 2D 
and find a similar enhancement of the inverted regions with respect  
to the 1D case. The corresponding expressions 
for $\vrt$ 
and $p_{min}^{\vrt}$ are listed in Tables \ref{vp2d} and \ref{pminn2d}.
The values of $p_{min}^{\vrt}$ are very close to those for 
$p_{min}^{\rt}$, with the exception of the $2\times 2$ lattice, for
which one respectively has $p_{min}^{\vrt}\approx 0.1847$ and 
$p_{min}^{\rt}\approx 0.1042$.

\subsection{Confining boundary conditions}

In order to complete the picture of the confining case, 
we have investigated the reaction efficiency for two cases, namely 
the $3\times 3$ and the $5 \times 5$ lattice. 

Let us first discuss the $3\times 3$  lattice. There are 
12 symmetry-distinct states altogether, i.e., 5 states with 
the heavy walker at a corner site CS: 

\[
\begin{array}{ccccccccc}
\mbox{State 1} &\qquad &\mbox{State 2}  &\qquad & \mbox{State 3} & \qquad &
\mbox{State 4} &\quad & \mbox{State 5}\\
&&&&&&&&\\
2-1-0 &\qquad & 2-0-0 &\qquad  & 2-0-1 &\qquad  & 2-0-0  &\qquad  & 2-0-0\\
|\hspace{0.6cm} |\hspace{0.6cm} | && |\hspace{0.6cm} |\hspace{0.6cm} | &&  
|\hspace{0.6cm} | \hspace{0.6cm} |  && |\hspace{0.6cm} |\hspace{0.6cm} | &&  
|\hspace{0.6cm} | \hspace{0.6cm} |   \\
0-0-0 & \qquad & 0-1-0 &\qquad & 0-0-0 & \qquad & 0-0-1 &\qquad & 0-0-0 \\
|\hspace{0.6cm}  | \hspace{0.6cm} | &\qquad &  |\hspace{0.6cm}  
|\hspace{0.6cm} | &\qquad &  |\hspace{0.6cm}  |\hspace{0.6cm}|
 && |\hspace{0.6cm} |\hspace{0.6cm} | &&  
|\hspace{0.6cm} | \hspace{0.6cm} | \\
0-0-0 &\qquad & 0-0-0 &\qquad & 0-0-0&\qquad & 0-0-0 &\qquad & 0-0-1 
\end{array}
\]

According to the numbering used in Fig. \ref{fig0} (right), the above
states would correspond to those for which $x_2=1$.
There are 5 more states in 
which the heavy walker is positioned on a midpoint boundary site MBS. These
states with $x_2=2$ can be pictorially represented as follows:

\[
\begin{array}{ccccccccc}
\mbox{State 6} &\qquad &\mbox{State 7}  &\qquad & \mbox{State 8} & \qquad &
\mbox{State 9} &\quad & \mbox{State 10}\\
&&&&&&&&\\
1-2-0 &\qquad & 0-2-0 &\qquad  & 0-2-0 &\qquad  & 0-2-0  &\qquad  & 0-2-0\\
|\hspace{0.6cm} |\hspace{0.6cm} | && |\hspace{0.6cm} |\hspace{0.6cm} | &&  
|\hspace{0.6cm} | \hspace{0.6cm} |  && |\hspace{0.6cm} |\hspace{0.6cm} | &&  
|\hspace{0.6cm} | \hspace{0.6cm} |   \\
0-0-0 & \qquad & 1-0-0 &\qquad & 0-1-0 & \qquad & 0-0-0 &\qquad & 0-0-0 \\
|\hspace{0.6cm}  | \hspace{0.6cm} | &\qquad &  |\hspace{0.6cm}  
|\hspace{0.6cm} | &\qquad &  |\hspace{0.6cm}  |\hspace{0.6cm}|
 && |\hspace{0.6cm} |\hspace{0.6cm} | &&  
|\hspace{0.6cm} | \hspace{0.6cm} | \\
0-0-0 &\qquad & 0-0-0 &\qquad & 0-0-0&\qquad & 1-0-0 &\qquad & 0-1-0 
\end{array}
\]
Finally, there are 
2 more states in which the heavy walker sits in the centrosymmetric 
site CES (i.e., one has $x_2=5$), namely

\[
\begin{array}{ccc}
\mbox{State 11} &\qquad &\mbox{State 12} \\
&&\\
0-0-0 &\qquad & 0-0-0 \\
|\hspace{0.6cm} |\hspace{0.6cm} | && |\hspace{0.6cm} |\hspace{0.6cm} | \\
0-2-0 & \qquad & 0-2-0 \\
|\hspace{0.6cm} |\hspace{0.6cm} | && |\hspace{0.6cm} |\hspace{0.6cm} | \\
0-0-1 & \qquad & 0-1-0 
\end{array}
\]

\begin{figure}[htbp]
\includegraphics[width=8cm,height=8cm]{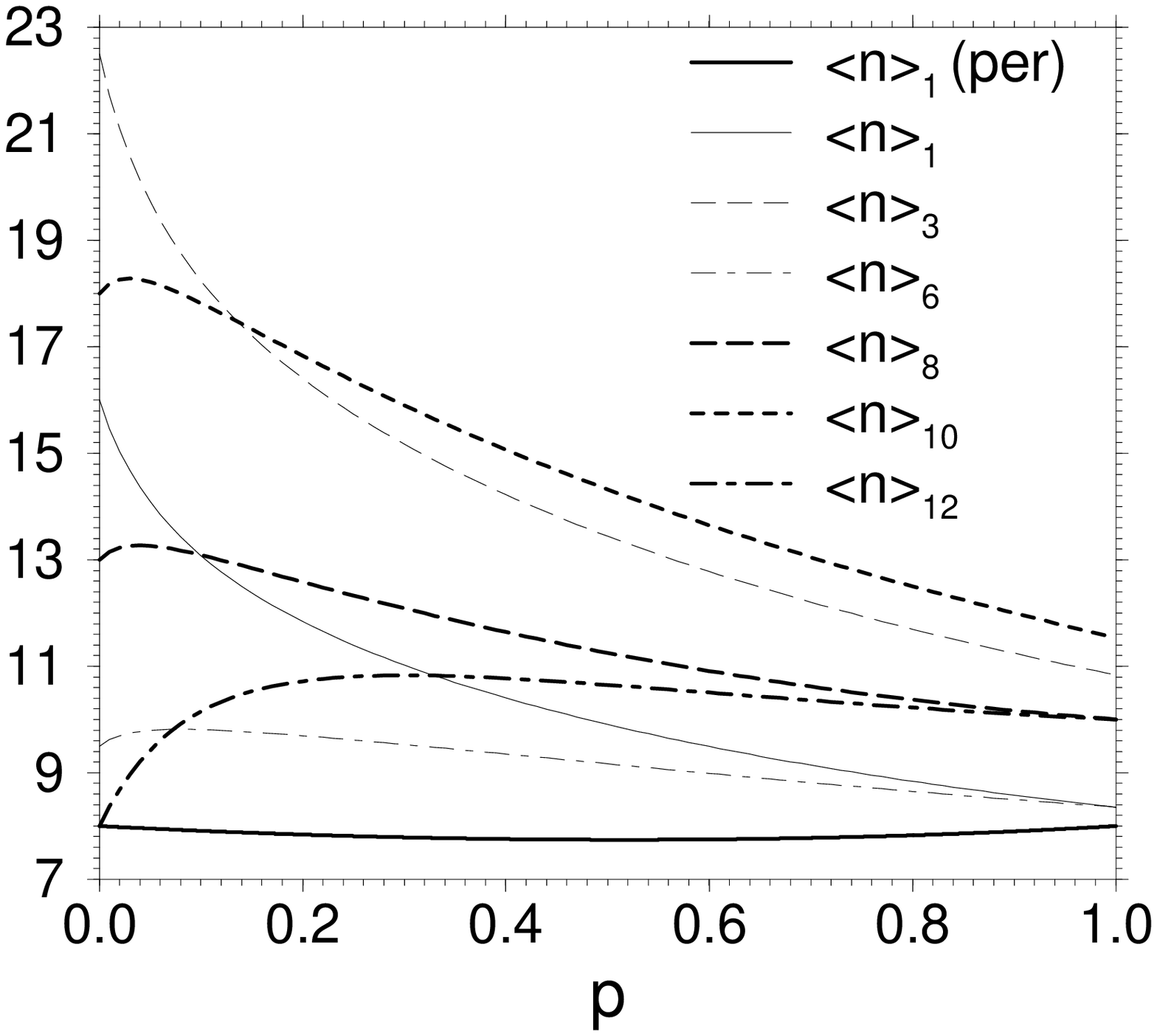}
\includegraphics[width=8cm,height=8cm]{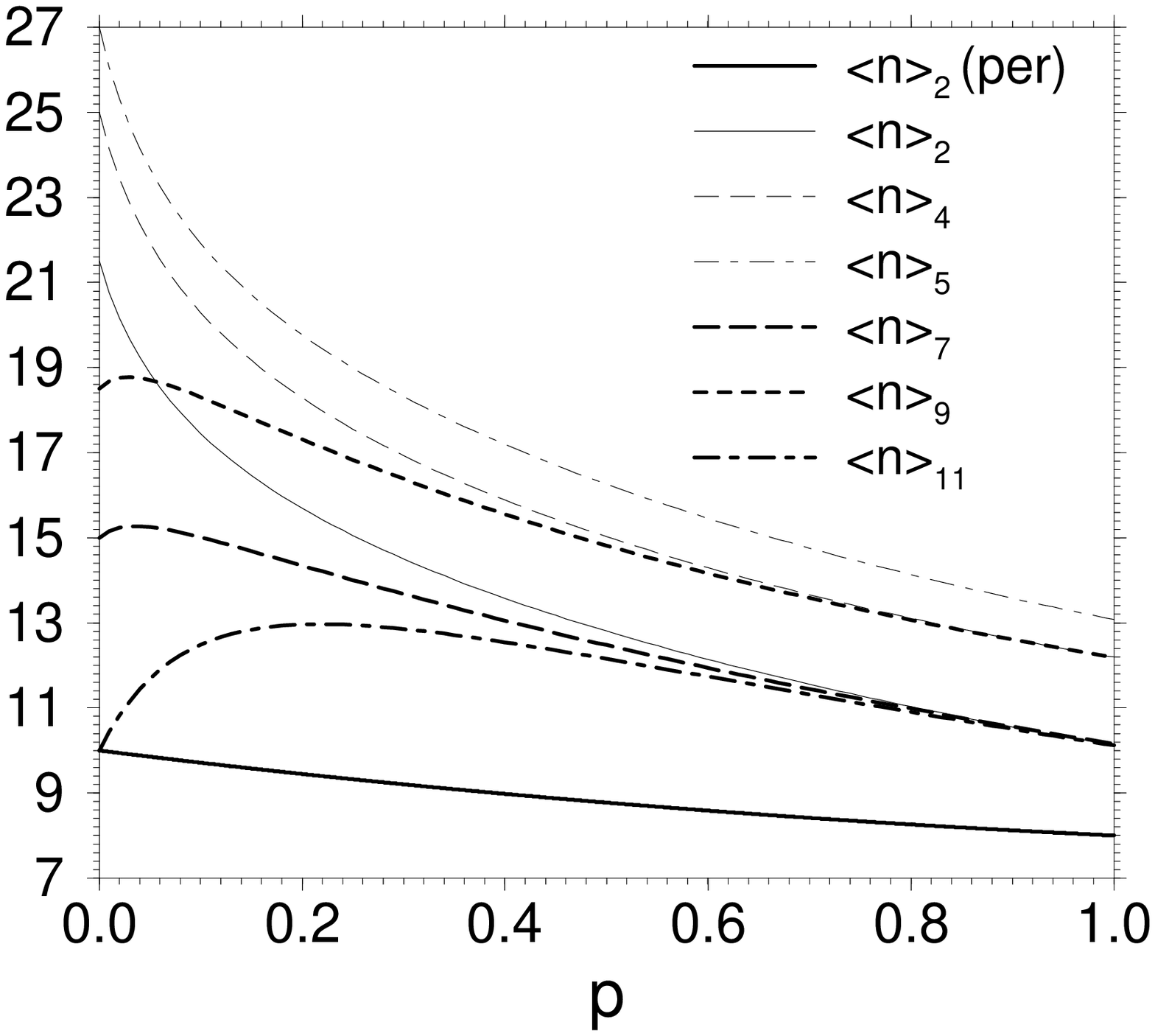}
\caption{\it \label{3x3lrt} $p$-behavior of the local reaction times
in the confining $3\times 3$ case. The initial states 1, 3, 6, 8, 10 and
12 in the confining case collapse to the same state when the boundary
sites are connected periodically, i.e., State 1 with 
contiguous walkers in the periodic lattice (see left graph). 
The initial states 2, 4, 5, 7, 9 and 11
also collapse to the same state, i.e., State 2 corresponding 
to walkers placed diagonally in the periodic lattice (see right
graph). The thick solid lines show the $p$-behavior in the periodic case.}
\end{figure}

The $p$ behavior of the local reaction times is depicted in Fig. 
\ref{3x3lrt}. The reaction times associated with the CS 
initial configurations 1-5 
turn out to be monotonically decreasing in $p$. 
In contrast, the MBS configurations 6-10 and the 
CES configurations 11-12 are associated with reaction times
displaying a single maximum.  
In the MBS case the maximum is located at relatively small values of
$p$, whereas in the CES case it is found at somewhat larger $p$ values.   
Interestingly, there are no initial conditions which give rise to a 
double inverted region, in contrast with the 1D case. These results are
also confirmed by numerical simulations (data not shown).

We now compare the above results with the periodic case. If one connects
the boundary sites of the confining lattice in a periodic fashion, one
finds that States 1, 3, 6, 8, 10 and 12 collapse into State 1,
i.e., the state in which the walkers are positioned aside of each other in
the periodic lattice. On the other hand, States
2, 4, 5, 7, 9 and 11 collapse into State 2, i.e., the state in which  
both walkers are positioned diagonal to each other in the periodic lattice. 
The plots corresponding to the reaction times in the periodic case 
are displayed in Fig. \ref{3x3lrt} and compared to the corresponding
plots for the reaction times in the confining case. For all values of $p$,
the local reaction times are smaller in the periodic case than in all
confining cases \footnote{with the exception of two initial configurations
for which in the $p=0$ limit the reactions times are 
found to be equal to those for the corresponding periodic cases}.
This behavior clearly differs from the 1D case, where 
for certain initial conditions and $p$-intervals the efficiency is larger 
in the confining case.  

Finally, we consider the behavior of the global averages. Consonant
with what was done in \ref{cobc}, one can define via equation (\ref{subglav}
)three global averages $\rt^{CS}=\rt^{x_2^0=1}$,$\rt^{MBS}=\rt^{x_2^0=2}$
and $\rt^{CS}=\rt^{x_2^0=5}$
for all three symmetry-distinct positions of
the heavy walker. In the present case, these averages are related to 
the local reaction times as follows:

\ba
\rt^{CS}&=&\frac{2\rt_1+\rt_2+2\rt_3+2\rt_4+\rt_5}{8},\nn \\
\rt^{MBS}&=&\frac{2\rt_6+2\rt_7+\rt_8+2\rt_9+\rt_{10}}{8} \nn \\
\rt^{CES}&=&\frac{\rt_{11}+\rt_{12}}{2}. \nn \\
\ea
The reaction time averaged over all initial conditions is 

\be
\label{aavrt}
\rt=\frac{4\rt^{CS}+4\rt^{MBS}+\rt^{CES}}{9}
\ee 

\begin{figure}[htbp]
\includegraphics[width=8cm,height=8cm]{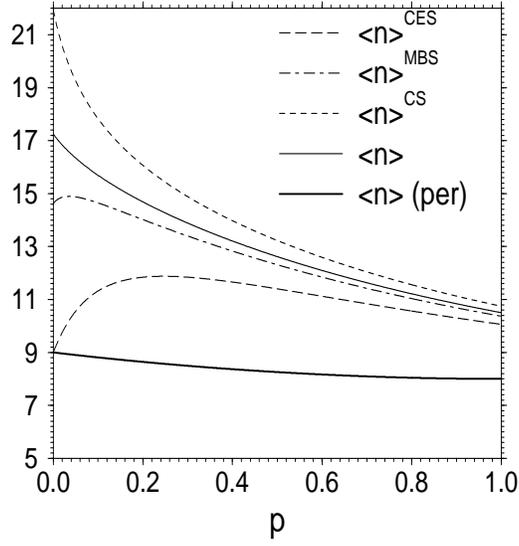}
\caption{\it \label{fig20} $p$-Behavior of the global reaction times 
in the confining $3\times 3$ case. The thick solid line corresponds to
the behavior of $\rt$ in the periodic $3\times 3$ case. }
\end{figure}

Fig. \ref{fig20} displays the $p$-behavior of the three global averages
and $\rt$. As in the 1D case, 
the efficiency of the reaction for a given $p$ 
increases as the heavy walker is moved away from the centrosymmetric site. 
On the other hand, one can see 
that $\rt^{CS}$ decreases monotonically with $p$, 
whereas $\rt^{MBS}$ and $\rt^{CES}$  
display a maximum. 
These maxima are averaged out when Eq. (\ref{aavrt}) is used to compute
$\rt$, leading as in the 1D case to monotonically decreasing behavior
of $\rt$. 
The global averages are all strictly larger than in the 2D periodic case, 
with the exception of $\rt^{CES}$ in the limit $p\to 0$
(see Fig. \ref{fig20}). 

We have also examined the behavior of the global averages in the $5\times
5$ case. In this case there are 6 global averages corresponding to 
the 6 possible symmetry-distinct initial positions of the heavy walker 
characterized by Euclidian distances 
$d_{CES}^0=0,1,\sqrt{2},2,\sqrt{5},\sqrt{8}$ from the centrosymmetric site. 
The results obtained are entirely consistent 
with the findings for the $3\times 3$ 
lattice, i.e., the efficiency increases when the heavy walker is moved
towards the interior of the lattice. 
For $d_{CES}^0\le\sqrt{2}$ the curves $\rt_i(p)$ display a a maximum
in $p$, whereas for larger $d_{CES}^0$ the decay is monotonic. 
Once again, the reaction time scaled by $\rt$ and computed by
averaging over all initial conditions, is monotonically decreasing.  

\section{General case}
\label{Sec5}

In this section we shall discuss how results for the periodic 
lattice case are modified if one allows for the possibility that 
walker 1 does not jump at a given time step. To 
this end, let us now assume that walker 1 and walker 2 respectively
jump with probabilities $p_1$ and $p_2$ at each tick of the clock. 
Results for this case can be easily obtained using the approach 
developed in \ref{marap}. Note that the initial scenario is 
recovered as a special case
if one sets $p_1\equiv 1$ and $p_2\equiv p$. 

\subsection{1D case}
\label{gen1d}

\begin{figure}[htbp]
\includegraphics[width=8cm,height=8cm]{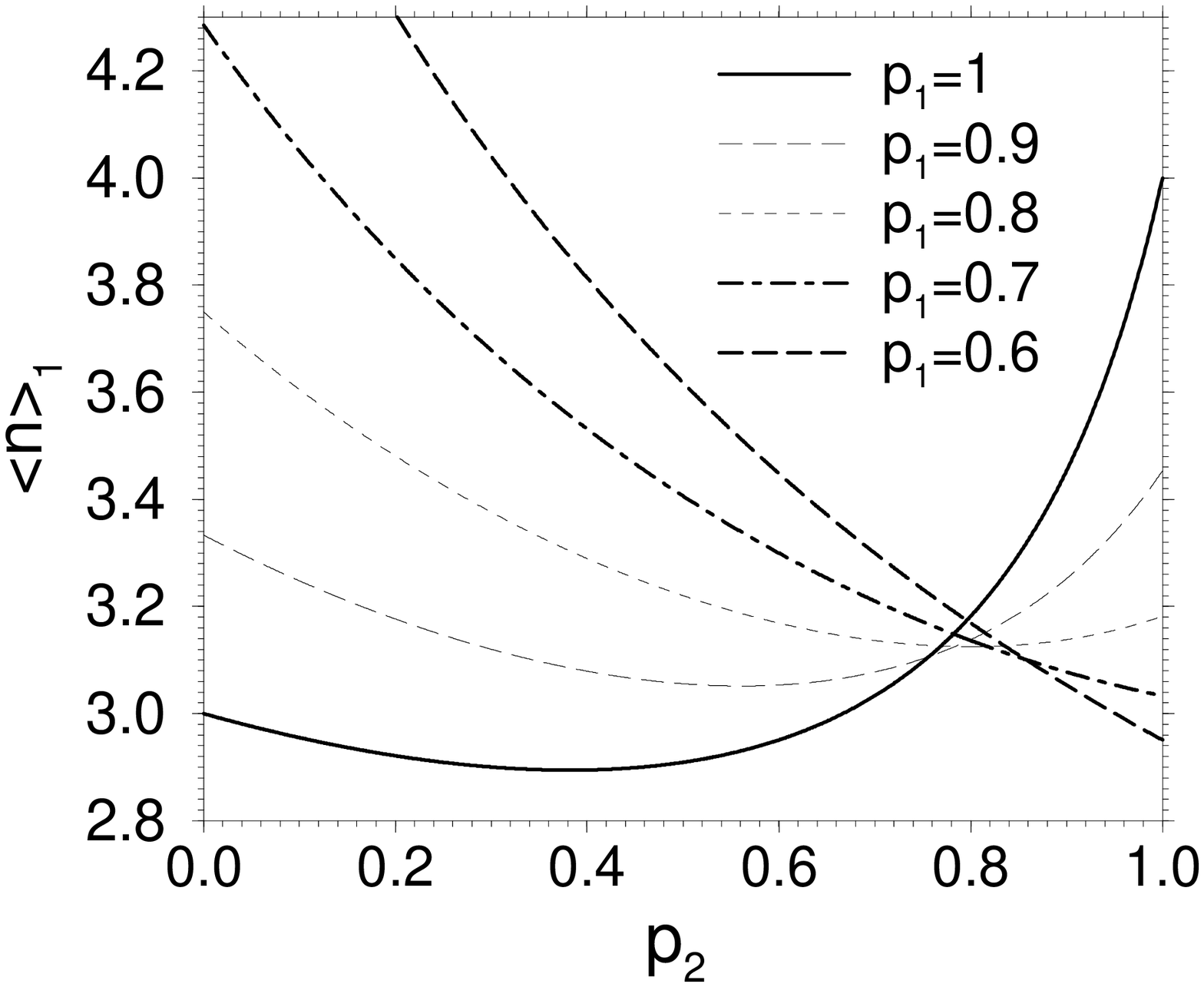}
\includegraphics[width=8cm,height=8cm]{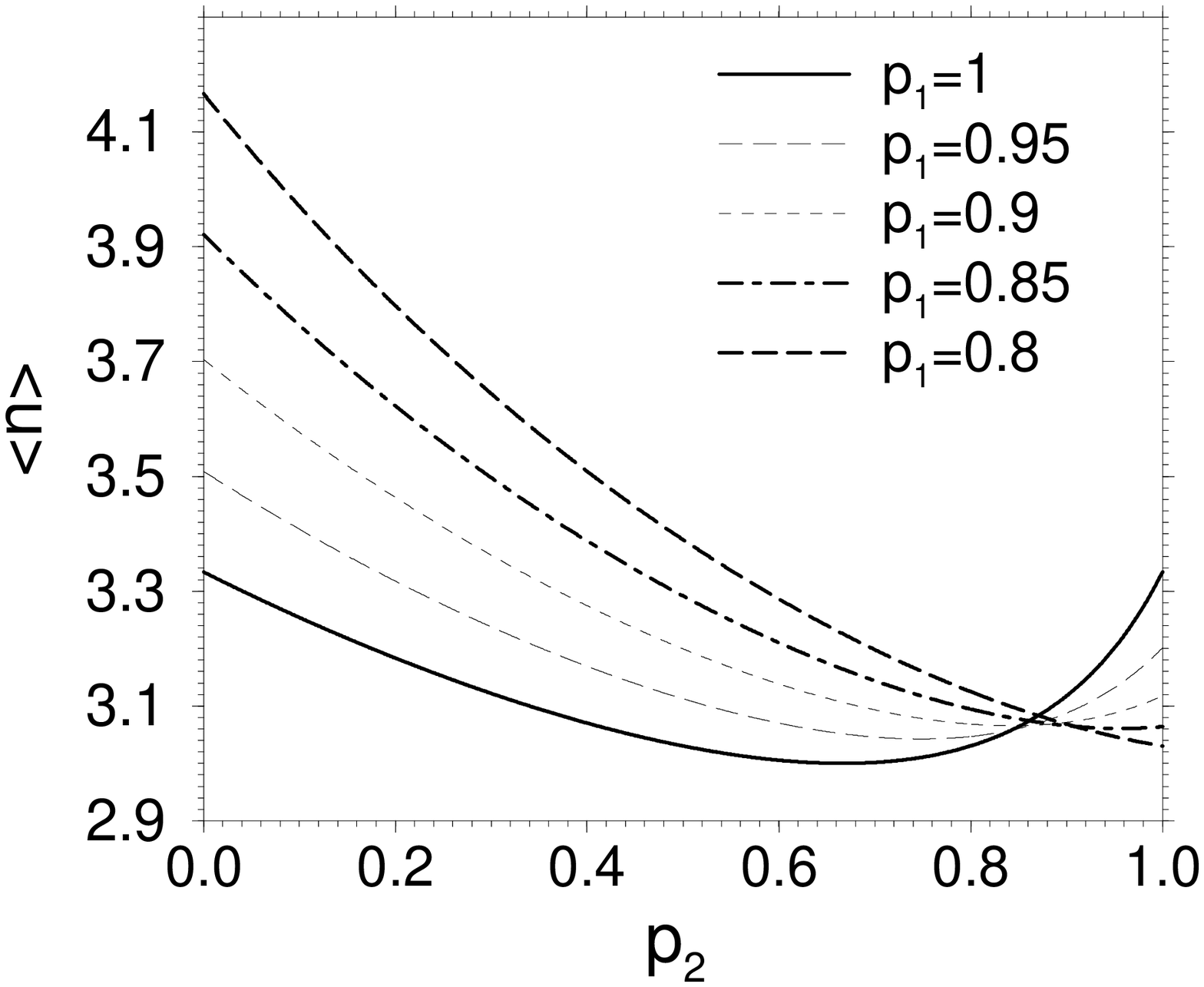}
\caption{\it \label{fig6} Behavior of the curves $\rt_1(p_2)$ (left)
and $\langle n\rangle(p_2)$ (right) for different values of $p_1$ in 
a periodic 1D lattice with $N=4$.}
\end{figure}

Let us first examine the behavior of the local reaction times 
$\rt_i(p_1,p_2)$ as a function of $p_2$ when the value of $p_1$ is
decreased below one. As our previous results for the $p_1=1$ case show, 
there is a monotonic decrease 
in the reaction time with respect to $p_2$ when
the state number $i$ takes even values (cf. Fig. \ref{fig2}) and
minima in $p_2$ for odd values of $i$.
Now, if one now gradually decreases $p_1$, the qualitative 
behavior of the $\rt_i(p_2)$ profiles does not change for even values 
of $i$, i.e. one still has $\frac{\partial \rt_i}{\partial p_2}<0$ 
for all values of $p_2$. 
In contrast, there is a qualitative change in behavior for even values of 
$i$; the inverted regions with $\frac{\partial \rt_i}{\partial p_2}>0$
vanish and one finds monotonically-decreasing profiles in this case
(see Fig. \ref{fig6}, left). The even-odd effect in $i$ is thus lost.   
  
Let us now examine the $p_2$ behavior of the global reaction time 
$\langle n\rangle$ when $p_1$ is decreased. Analytic 
expressions for $\rt$ as a function of $p_1$ and $p_2$ for periodic 
lattices of increasing size are given in Appendix A 
(see Table \ref{tabenct2}). These expressions
are symmetric with respect to exchange of $p_1$ and $p_2$.
Here, for odd values of $N$  
$\rt$ decreases monotonically with $p_2$ regardless of the value of 
$p_1$, i.e., $\frac{\partial \langle
n\rangle}{\partial p_2}<0$ for all values of $p_2$. 
In contrast, the behavior 
in the case of an even lattice again depends on the value of $p_1$. 
As we have seen, for $p_1=1$ one has an inverted region where
 $\frac{\partial \langle
n\rangle}{\partial p_2}>0$. However, if one gradually decreases 
$p_1$ the minimum of $\langle n\rangle(p_2)$
is shifted to the right and below a characteristic $N$-dependent 
threshold value $p_{1,c}^{\rt}$ the inverted region finally
vanishes and the curves $\rt(p_2)$ become monotonically 
decreasing. 

The critical values $p_{1,c}^{\rt}$ are displayed in Table \ref{p1c1d}
for periodic lattices of increasing size.
These values are obtained as follows.
Denote by $p_{2,min}^{\rt}$ the value of $p_2$ for 
which a minimum of $\rt$ as a function of $p_2$ is observed.
The value of $p_{2,min}^{\rt}$ as a function
of $p_1$ is then easily obtained from the condition 
$\frac{\partial \langle n\rangle}{\partial p_2}=0$ for $p_2=p_{2,min}^{\rt}$.
One has $\frac{\partial p_{2,min}^{\rt}}{\partial p_1}<0$,i.e.,
the probability $ p_{2,min}^{\rt}$ is shifted to larger values
with decreasing $p_1$ until it becomes equal to one. This 
condition yields a polynomial equation in $p_1$ with a single root in 
the interval [0,1]; this root corresponds to the threshold
value $p_{1,c}^{\rt}$ below which 
the profiles $\rt_i(p_2)$ decrease monotonically. 
Note that one has $p_{1,c}^{\rt} \to 1$ as $N\to\infty$. 

Fig. \ref{fig6} (right) shows the suppression of the inverted zone in the case
of the $N=4$ lattice. 
Several plots of $\langle n\rangle(p_2)$ corresponding to decreasing 
values of $p_1$ are displayed. It is seen that
the minimum of the curves shifts to the right with decreasing $p_1$
until it eventually vanishes for 
$p_1<p_{1,c}^{\rt}\approx 0.834$ (cf. Table \ref{p1c1d}). 
Interestingly, the reaction time decreases with $p_1$ for a fixed
value of $p_2$ in the region
where $p_2 \stackrel{<}{\sim} 1$, while the opposite is true 
for not too large values of $p_2$. This behavior is also confirmed by 
numerical simulations (data not shown).   

A similar suppresion of the inverted region with decreasing $p_1$
is also observed at the level of
the global variance $\vrt$. The corresponding polynomial expressions 
as well as the critical values $p_{1,c}^{\vrt}$ 
are listed in Tables \ref{tabenct3}, \ref{tabenct5} and
\ref{p1c1e}. As it turns out, the value of $p_{1,c}^{\vrt}$ 
is almost identical to $p_{1,c}^{\rt}$ (see 
Tables \ref{p1c1d} and \ref{p1c1e}). 

\subsection{2D case}

\begin{figure}[htbp]
\includegraphics[width=8cm,height=8cm]{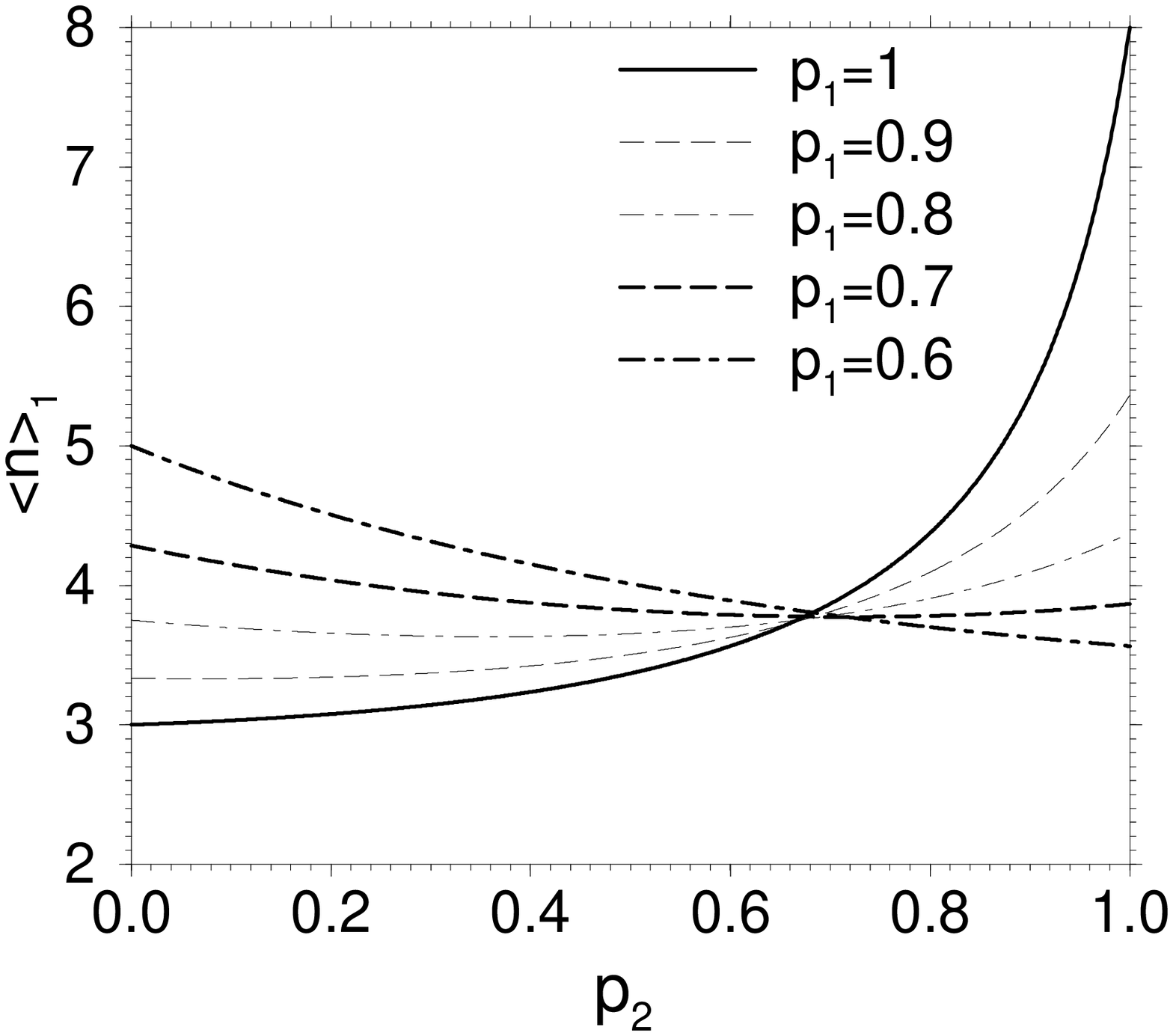}
\includegraphics[width=8cm,height=8cm]{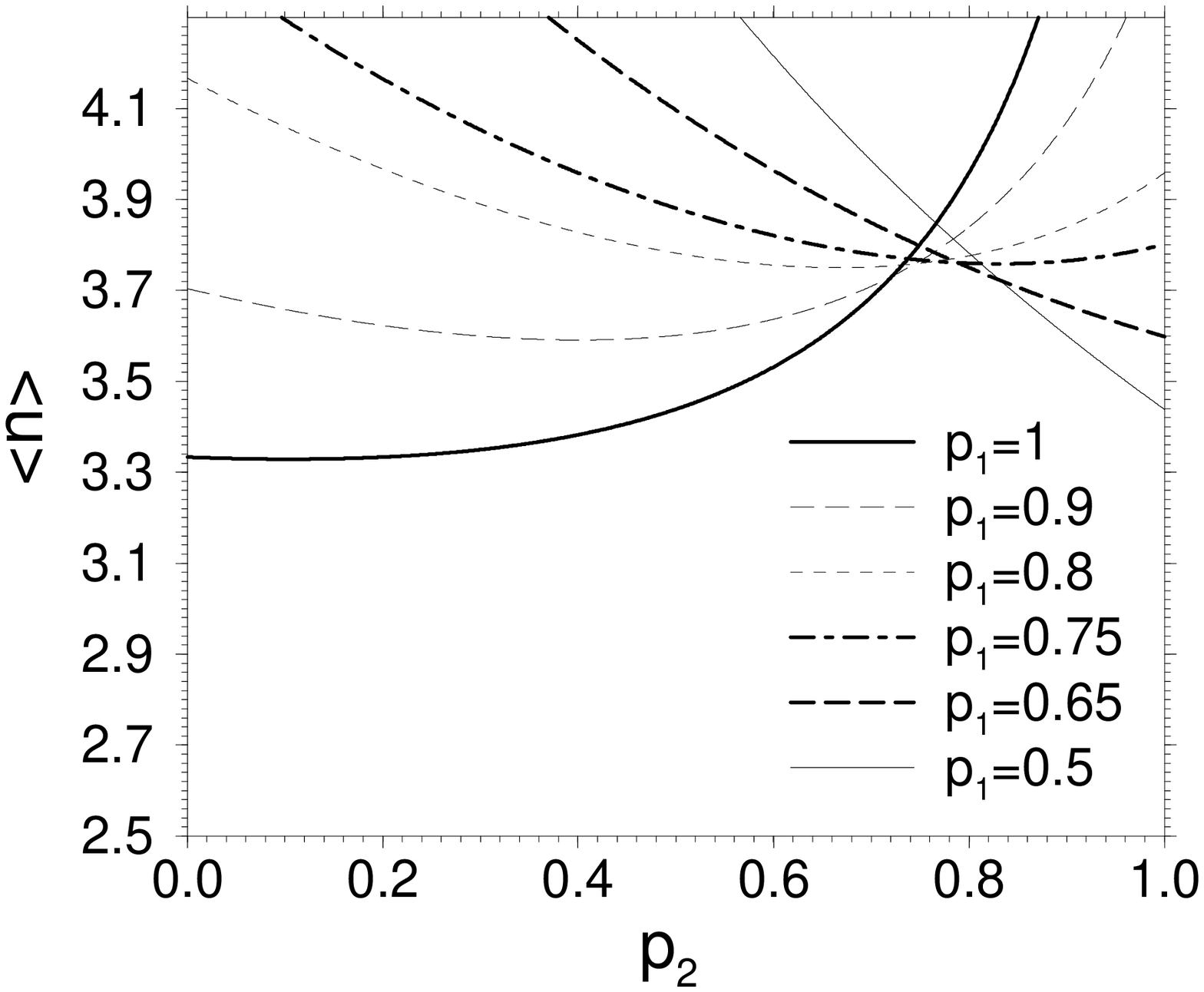}
\caption{\it \label{fig14} Behavior of the curves 
$\rt_1(p_2)$ (left) and $\rt(p_2)$ (right) for different 
values of $p_1$ in a $2\times 2$ periodic square planar lattice.}
\end{figure}

As in the 1D case, inverted $p_2$ regions are suppressed with decreasing
$p_1$, both at the level of the local and the global reaction times. 
Fig. \ref{fig14} displays the behavior of the 
$\langle n\rangle_1$ and $\rt$ as a function of $p_2$
for the $2\times 2$ lattice. 
In both cases, the curves become monotonically decreasing 
below a given threshold value of $p_1$. In contrast, the curves 
$\langle n\rangle_2(p_2)$ remain monotonically decreasing for any
values of $p_1$. The $p$-behavior of $\rt$ is thus qualitatively 
similar to the 1D case (cf Figs. \ref{fig6} and \ref{fig14}). 
The analytic expressions 
for $\langle n\rangle$ in terms of $p_1$ and 
$p_2$ as well as values 
of $p_{1,c}^{\rt}$ for even square lattices of increasing size
are listed in Appendix B (Tables \ref{np1p22d} and
Table \ref{p12d}). 

For 2D lattices the limiting behavior $p_{1,c}^{\rt} \to 1$ as 
$N\to \infty$ is the same as for 1D. Here, however, 
for a given number $N$ of lattice sites the
values of $p_{1,c}^{\rt}$ displayed in Table \ref{p12d}
are systematically smaller than those shown in Table 
\ref{p1c1d}. This fact suggests that in higher 
dimensions the inversion effect is more robust
vis a vis a decrease in the mobility of walker 1. 
Finally, suppression of the inverted region with decreasing 
$p_1$ is also seen in the behavior of the global variance 
$\langle v \rangle$. The polynomial expressions for $\vrt$ and the
associated values of $p_{1,c}^{\vrt}$ are displayed in
Tables \ref{vp1p22d} and \ref{pminv2d}. 

\section{Discussion}
\label{Sec6}
 
Let us now summarize the main results obtained so far and
introduce some physical arguments to explain them. 

\subsection{Results for periodic lattices}

The existence of an inverted region in the case of 
a 1D periodic lattice is governed by a parity effect. For even
values of the interparticle distance $d^0$ no inverted regions are observed. 
However, for odd
values of $d^0$  the reaction time as a function of $p$ displays a single
minimum for a $p$-value larger than $1/2$, which rapidly shifts towards one
with increasing $d^0$ and increasing lattice size. In 2D, a similar even-odd
effect holds only for sufficiently small lattices; for larger lattices 
exceptions to this rule become increasingly frequent. On the other hand, 
the inverted regions become broader with respect to the 1D case and the 
enhancement of the reaction's efficiency within the inverted region 
also becomes much larger (cf. Table \ref{compdelta}). 
 
As already anticipated in \ref{subper}, the above even-odd effect in 1D 
is closely related to the fact that in
prereactive configurations where both walkers are located at contiguous sites
an asynchronous step leads to reaction with a probability twice as large
as for a synchronous step (1/2 vs. 1/4). For such a configuration asynchronous 
dynamics due to a lower system temperature
minimizes the probability of mutual avoidance once the particles are within 
the interaction zone (in which the next time step may lead to reaction). 
For sufficiently high values of $p$ the number of 
statistical paths leading to such nearest-neighbor
configurations are higher for odd values of $d^0$ than
for even values. In this high $p$ regime, 
the reaction time increases with $p$ for odd $d^0$, while it 
decreases with $p$ for even values of $d^0$. 
The number of initial conditions with odd values of $d^0$ is larger 
for an even value than for an odd value of $N$; 
thus the total increase in $p$ of the 
associated reaction times in the former case
overcomes the $p$-decrease of the reaction times associated with 
even-valued initial distances and the global reaction time $\rt$ increases. 

In 1D, the size of the inverted region for odd values of the interparticle 
distance $d^0$ must decrease
with increasing $d^0$ (Fig. \ref{fig2}) or increasing lattice size $N$
(Fig. \ref{fig1}), since discrete diffusion over long distances 
becomes the rate limiting step. 
In 2D, the enhancement of the even-odd effect with respect to the 1D case
can be understood as follows. For a given number of sites $N$, the
initial-condition-averaged distance between the particles is smaller 
than in 1D, thereby
reducing the role of long-distance hopping with respect to
mutual avoidance in the interaction zone. In addition, the
probability of reaction for nearest-neighbor configurations in 
a sufficiently large lattice is four times higher for an asynchronous 
than for a synchronous step (1/4 vs. 1/16).    

Finally, we have seen that the inverted regions in the
periodic case are suppressed if the mobility of the light walker lies below
a size-dependent threshold value. This is reasonable, since 
a prerequisite for the onset of the even-odd effect is that both particles
must first have a minimal mobility to come into the vicinity of each other.
If this is not the case, the dynamics is controlled by the relative
diffusion coefficient, which is a monotonically increasing function 
of $p_1$ and $p_2$. 

\subsection{Results for confining lattices}

In small confining lattices the behavior of the reaction time is even 
more complex. In 1D, the reaction
time may decrease monotonically, increase monotonically, display
a single maximum or even a maximum and a minimum, giving rise to inverted
regions in the last three cases (cf. Fig. \ref{fig17}). 
Those initial conditions with the heavy
walker at a boundary site do not give rise to inverted regions, whereas
all the others do. Initial configurations where the light walker is on a 
boundary site and the heavy walker is next to it result in reaction times which
increase monotonically with $p$, thereby giving rise to an inverted region
of maximum size. In particular, the $p=0$ case (one walker plus trap case)
then becomes more efficient than the $p=1$ case (case of identical walkers).
Double inverted regions are observed for those initial
states where both walkers are at interior sites 
next to each other with the light walker closer to the lattice boundary 
or at the same distance from the
boundary as the heavy walker. 
In contrast, no double inverted regions are observed in the 2D cases studied. 

Both in the 1D and the 2D case, the global reaction time $\rt^{x_2^0=i}$
(average over the positions of the light walker)
displays a local maximum as a function of $p$ as long as the heavy walker
starts sufficiently close to the geometric center of the lattice, 
and it decays monotonically with $p$ as the heavy walker is shifted
away from the center (cf. Figs. \ref{globeh} and \ref{fig20}). 
The reaction time $\rt$ is found 
to decrease monotonically with increasing 
$p$, in contrast with the case of a periodic lattice with 
an even number of sites.

\begin{figure}[htbp]
\includegraphics[width=9cm,height=4cm]{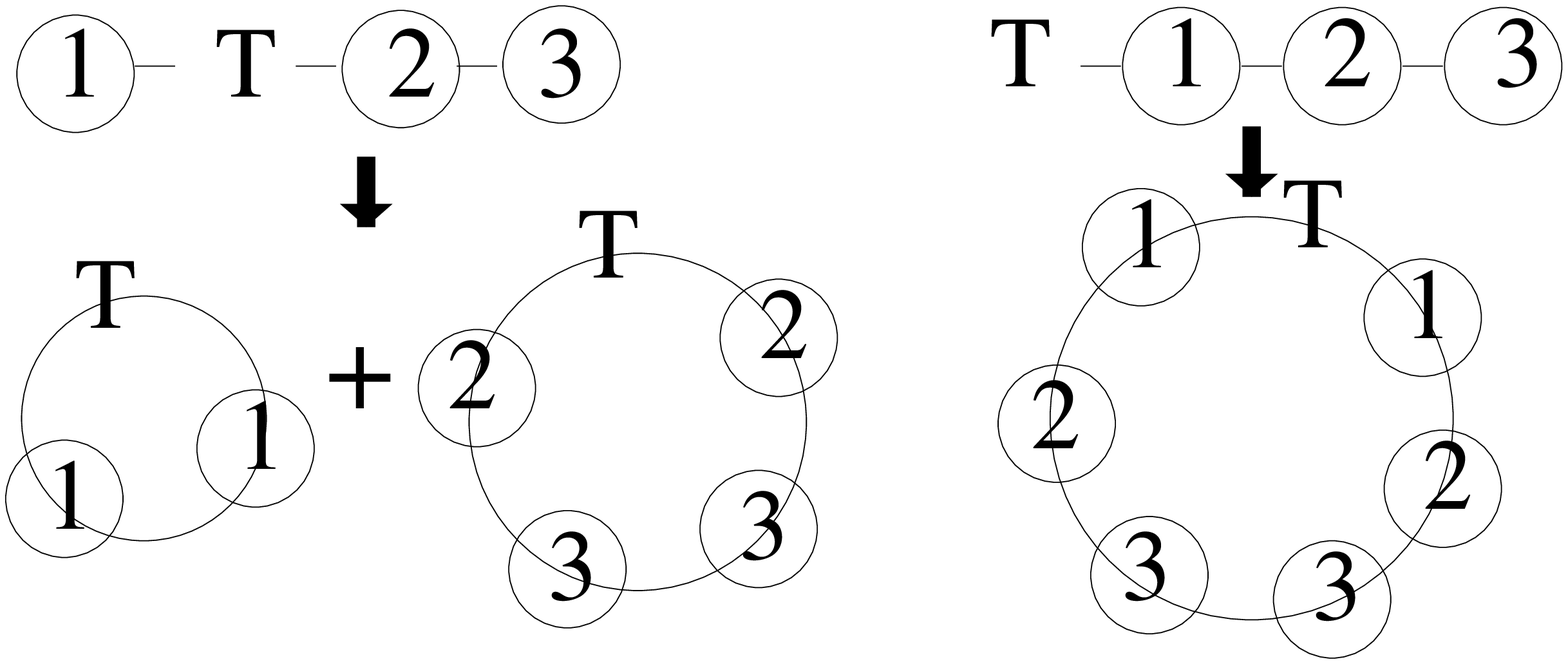}
\caption{\it \label{equivbc} Correspondence between a confining four-site 
1D system with a deep trap $T$
and periodic systems of different sizes for the two possible
symmetry-distinct trap configurations. Sites with same numbers 
are symmetry-equivalent. In our model, the heavy walker 2 plays
the role of the deep trap T in the figure. For $p=0$ 
the system in the left subfigure will be equivalent 
to a three- (five-) site system if walker 1
starts to the left (right) of walker 2. Similarly, 
the system in the right subfigure will be equivalent to a 
seven-site periodic system. From this picture one can easily infer 
that the motion of the trap in the case $p>0$ 
is associated with lattice size fluctuations
in the periodic lattice representation.}  
\end{figure}

This complex behavior for confining boundary conditions 
can, at least in 1D, be understood in terms of an 
description based on the concept of a ``fluctuating'' lattice.
To illustrate the use of such a metaphor, consider the 1D case subject 
to the choice $p=0$. Here, walker 2 plays the role of an 
immobile trap which divides the
lattice into two disconnected sublattices. Depending on its initial position,
walker 1 will perform a random walk between the reflecting and the 
absorbing boundary of one of these sublattices. 
From the standpoint of the reaction time (see Fig.
\ref{equivbc}), the motion of walker 1 in each sublattice is equivalent 
to its motion in a periodic lattice. For most
initial conditions, the size of the corresponding periodic lattice is
larger than the size of the original confining lattice, implying that 
for a given value of $N$ the reaction time 
(a monotonically increasing function of $N$) also becomes larger than in 
the periodic case. Thus, in the small
$p$ limit the reaction is less efficient in the confining
case than in the periodic case. An exception is found for the $N(N-1)/2$-th
initial condition (see e.g. the 
curve for $\rt_6$ in Fig. \ref{confvsper}, right), 
for which the effective size
of the associated periodic lattice is smaller than $N$. Such configurations
are no longer found in the 2D cases we have examined, where
periodic lattices are more efficient than confining ones, with the
exception of two cases in the limit $p\rightarrow 0$ 
for which the efficiency becomes the same (cf. Fig. \ref{3x3lrt}). 
This behavior is due to the fact that the trapping effect exerted by the 
heavy walker and the confining lattice boundary on the light walker is 
lost in higher dimensions. 

If one now slowly increases $p$, the above effective description in terms 
of a periodic lattice remains valid, since walker 2 remains a deep
trap. However, the fact that walker 2 can now move leads to time fluctuations
of the lattice size of the associated periodic lattice. In the case where 
walker 2 starts off from a boundary site (corresponding to the first $N$ 
initial conditions), fluctuations can only lead to a smaller effective size
than the initial one, resulting in a monotonic decrease of the reaction 
time with respect to the $p=0$ case. However, for any other initial condition
with walker 2 at an interior site small fluctuations in the position of
the trap can in many cases lead to a larger effective size, resulting 
in an increase of the reaction time. On the other hand, for somewhat larger 
values of $p$ 
excursions of walker 2 towards the boundaries become more frequent, 
resulting in a more efficient confinement of walker 1 and thereby in a
decrease of the reaction time. In other words, the effect due to lattice 
size fluctuations is overcome by the significant increase in the 
relative diffusion coefficient,
which is a monotonically increasing function of $p$ \cite{jon1}. 
The reaction time,
which in a continuum description is inversely proportional to the
diffusion coefficient, will therefore decrease with $p$ in this regime. 
As a result of the combination of the above low-$p$ and high-$p$ effects, 
for those configurations where walker 2 is sufficiently far from the
boundary and the initial particle separation is sufficiently large (such 
that the continuum picture applies for sufficiently
large $p$) the reaction time first increases and then decreases with $p$. 

The additional increase of the reaction time with $p$ in the 
high $p$ region for the subset of initial conditions associated with 
a double inverted region can possibly be explained in terms of an 
enhancement of mutual avoidance effects when particles are close to each 
other and far from the boundary \footnote{Recall that double inverted regions
are associated with initial conditions which fulfil these two conditions.}, 
resulting in an increased escape probability. 
The absence of this additional high-$p$ inverted region 
in the 2D case is probably due to the existence of 
additional degrees of freedom which
no longer make a description in terms of the periodic case possible. 
One could therefore expect that the even-odd effect leading to the 
high-$p$ inverted region does no longer hold and 
that the dynamics is essentially governed by long-range 
diffusion, resulting in a $p$-decrease of the reaction time. 

In 1D, the disappearance of the inverted region associated with the 
the global encounter time $\rt^{x_2^0=i}$
as walker 2 is shifted away from the inmost lattice sites to
the boundaries could be explained
by the fact that lattice size fluctuations responsible for the low
$p$ inverted region play a less important role when the initial 
distance between the walkers becomes larger. 
Clearly, the average of the interparticle 
distance over the positions of walker 1 becomes larger when walker 2 
is shifted away from the centrosymmetric site. 

Finally, the monotonic decrease of the global reaction time $\rt$ 
over the whole $p$ range can be explained as follows. In the small
$p$ limit the effective shortening of the lattice through the motion of
walker 2 from the boundary to the bulk for the first $N$ initial
conditions overcomes the effective increase in system size for
the other initial conditions. 
In the opposite limit of large $p$, the even-odd effect
leading to mutual avoidance in the effective periodic-lattice description
is weakened due to the lattice size fluctuations, and the 
global reaction time decreases.  

\section{Conclusions}
\label{Sec7}

The principal conclusion that can be drawn from the study presented in
this paper is that, in characterizing the
efficiency of  (two-channel)  reactions between independently-mobile
reactants on a lattice, for small systems one may find regimes where,
counter one's intuition, the reaction efficiency can decrease with
increasing "temperature."   Moreover, both minima and maxima in the
reaction efficiency have been found and quantified, thus establishing
the existence of one or more "inverted regions."  We stress that, in
contrast to Marcus theory, the emergence of these anomalous regions is a
consequence of a (strictly) classical interplay between system geometry
and individual particle dynamics (and the importance of interactions
within a certain reaction radius or "zone").  We have demonstrated and
quantified that these inverted regions arise in 1D and 2D
lattices for a wide class of initial conditions. Also, we note that
numerical evidence has been presented in \cite{eabad}
that the effect in present in 3D, and for cases where the reaction
"rules" are different from the ones adopted in this work \cite{eabad2}.

So that the plethora of results reported here not obscure a principal
insight, we emphasize that the discrete character of the geometric
support is an important (perhaps critical) ingredient in generating the
emergence of an inverted region.  It is precisely the discrete geometry
of the lattice which allows one to distinguish between reaction channels
(viz., irreversible termination of the reaction owing to same site
occupancy or nearest-neighbor crossing).  Such distinctions are lost in
the continuum limit, where the efficiency of
a diffusion-reaction process is characterized universally by an 
increase in reaction efficiency with respect to an
increase in system temperature.

In the cases we have studied here, much of the observed anomalous 
behavior can be understood in terms of a competition 
between diffusional transport over long distances and the probability 
of mutual avoidance inside the interaction zone. A very large jump 
probability $p$ may favor transport over long distances, since it increases 
the relative diffusion coefficient. However, it may also decrease the 
residence time
in the interaction zone. As we have seen, the latter effect may become 
dominant in sufficiently small lattices and thus determine the anomalous
decrease of the efficiency with temperature. Remarkably, the 
fluctuation-driven resonance and antiresonance phenomena reported in this
work do not involve coloured noise, in contrast with other 
first-passage problems such as resonant activation \cite{doerra}. 

The study presented here was motivated to a large extent by the
remarkable and unexpected results found in
studying experimentally systems at the nanolength scale.  It is
premature to suggest that the effects uncovered
here might be found in a particular experiment, but it is certainly not
premature to note that the sophistication
of these experiments has now reached a point where the presence of an
"inversion regime" might be explored.

From a theoretical point of view, there are many facets of the present
problem that need to be developed. For example, one can study
the effect of modifying the reaction rules. In this context, the case
of a reaction probability $p_R<1$ per collision is particularly 
interesting. An enhancement of the anomalous behavior can
be expected in this case, since the
efficiency of recollisions in the interaction zone can be strongly
diminished by amplifying the simultaneous
motion of the two reactants. Such 
counterintuitive effects are rather common in binary reaction systems. 
For example, Saxton has recently considered a binary reaction with 
$p_R<1$ between two simultaneously moving walkers 
in a percolation cluster with blocking sites \cite{saxt}. This 
model was recently proposed to explain protein motion in lipid rafts. Saxton 
came to the interesting conclusion that at sufficiently low reaction rates 
increasing the obstacle concentration might result in an anomalous 
decrease of the reaction time due to an enhancement of recollisions 
in the interaction zone. Similar effects are currently
being investigated for binary reactions in other types of disordered
media, e.g. so-called small-world networks, in which a very small 
number of long-range connections between sites
may have drastic effects upon the reaction time.

Finally, since non-nearest neighbor jumps \cite{jon2} are likely to become 
of increasing importance with an increase in the system temperature, 
it is of some interest to incorporate this facet into the present model. 
Moving from a scenario in which one analyzes the problem in terms of a 
symmetric random walk, incorporating "drift" and accounting for 
deactivation by regular or periodic impurities in the
system opens up connections to an even broader range of experimental 
phenomena.  It may be hoped that the new insights arising from the study 
of the above systems along with non-classical approaches will help to 
improve the current understanding of encounter-controlled processes in 
small-size systems. 

\section{Acknowledgments}

We thank J. L. Bentz for a comprehensive numerical check of our results
for the $3\times 3$ confining lattice and the general 1D case treated in
\ref{gen1d} and Prof. G. Nicolis for helpful discussions.

\newpage

\section*{Tables in the main body}

\begin{table*}[htbp]
\begin{ruledtabular}
\begin{tabular}{cc}
$N$ & $\langle n \rangle $ \\
\hline
2  & $ 2/(2-p)$ \\
&\\
3 & 2 \\
&\\
4 & $(10/3) (3p-4)/(p^2+2p-4)$ \\
&\\
5 & $4 (2p-5)/( p^2-4)$ \\
&\\
6 & $(28/5) (p^2-10p+10)/(p^3-4p^2-4p+8) $ \\
&\\
7 & $(4/3) (p^2+8p-14)/(p^2-2)$ \\
&\\
8 & $(12/7) (13p^3+6p^2-126p+112)/((p-2)(p^3+6p^2-8))$ \\
&\\
9 & $10 (2p^3-5p^2-16p+24)/((p^2+2p-4)(p^2-2p-4))$ \\
&\\
10 & $(22/9) (7p^4-76p^3+16p^2+288p-240)/(p^5-6p^4-12p^3+32p^2
+16p-32)$ \\
\end{tabular}
\end{ruledtabular}
\caption{\label{tabenct} Analytic expressions for $\rt$ 
(1D periodic lattices of increasing size $N$)}.
\end{table*}

\begin{table}[htbp]
\begin{ruledtabular}
\begin{tabular}{c}
\hspace{-.2cm}$N$\hspace{3.5cm}$p_{min}^{\rt}$ \\
\hline
 2 \hspace{3cm}    .000000 \\
 4  \hspace{3cm}  .666667   \\
 6  \hspace{3cm} .859648 \\
 8  \hspace{3cm} .920393 \\
 10 \hspace{3cm} .948308 \\
 12 \hspace{3cm} .963611 \\
 14 \hspace{3cm} .972951\\
 16 \hspace{3cm} .979086 \\
 18 \hspace{3cm} .983337 \\
20 \hspace{3cm} .986407\\ 
22 \hspace{3cm} .988698\\
\hspace{-.2cm}24 \hspace{3cm} .99045\\
\hspace{-.2cm}26 \hspace{3cm} .99183\\
\hspace{-.2cm}28 \hspace{3cm} .99292\\
\hspace{-.2cm}30 \hspace{3cm} .99381\\
\hspace{-.2cm}32 \hspace{3cm} .99455\\
\hspace{-.2cm}34 \hspace{3cm} .99515\\
\hspace{-.2cm}36 \hspace{3cm} .99567\\
\end{tabular}
\end{ruledtabular}
\caption{\label{pmintab} Values of $p_{min}^{\rt}$ 
(1D periodic lattices of increasing size $N$)}.
\end{table}

\begin{table}[ht]
\begin{ruledtabular}
\begin{tabular}{cc}
$N$ & $\langle v \rangle $ \\
\hline
2 & $ 2p/(2-p)^2$ \\
&\\
3 & $2$ \\
&\\
4 & $(2/3)(192-316p+152p^2-15p^3)/(p^2+2p-4)^2$ \\
&\\
5 & $4(-2p^3+25p^2-80p+84)/(p^2-4)^2$ \\
&\\
6 & $(28/5)(-1096p+768p^2-206p^3+28p^4-p^5+512)/(8-4p+p^3-4p^2)^2 $ \\
&\\
7 & $(4/3)(-320p+88p^2+8p^3+p^4+252)/(p^2-2)^2$ \\
&\\
8 & $(12/7)(-13p^7+252p^6+546p^5-4736p^4-2184p^3$\\
& $+35520p^2-50848p+21504)
/((p^3+6p^2-8)^2(p-2)^2)$\\
&\\
9 & $2(-10p^7+253p^6-816p^5-2748p^4+11872p^3+5552p^2$\\
&$-42496p+29568)
/((p^2-2p-4)^2(p^2+2p-4)^2)$ \\ 
\end{tabular}
\end{ruledtabular}
\caption{\it \label{tabvar} Analytic expressions for 
$\langle v \rangle$ (1D periodic lattices of increasing size $N$)}
\end{table}

\begin{table}[htbp]
\brt
\begin{tabular}{cc}
$N$ &\hspace{3cm} $p_{min}^{\vrt}$ \\
\hl
2  &\hspace{3cm} .000000\\
4  &\hspace{3cm} .666667\\
6  & \hspace{3cm}.852189\\
8  & \hspace{3cm}.914315\\
10 & \hspace{3cm}.943742\\
\end{tabular}
\ert
\caption{\label{pmintab2} Values of $p_{min}^{\vrt}$ 
(1D periodic lattices of increasing size $N$)}
\end{table}

\begin{table}[htbp]
\brt
\begin{tabular}{cc}
$N$ & $\langle n \rangle $ \\
\hl
4& $(2/3)(40-31p)/(-p^2-6p+8)$\\
&\\
9& $8(9-2p)/(8-p^2)$\\
&\\
16& $(8/45)(1191p^3+252p^2-14252p+13184)/(3p^4+24p^3-56p^2-96p+128)$\\
&\\
25& $(76/3)(-9p^3+2p^4+320-124p^2-48p)/(p^5+4p^4-44p^3-144p^2+64p+256)$\\
&\\
36& $(2/1225)(169521024p^3-99291648p^2-21958740p^5+760284p^6$\\
&$+247652352-296662016p+494505p^7+614144p^4)/(5376p^3$\\
&$-8704p^2-1176p^5-256p^6-6144p+8192+54p^7+5p^8+2688p^4)$\\
\end{tabular}
\ert
\caption{\label{n2d} Polynomial expressions for 
$\langle n \rangle$ (2D periodic square planar lattices of increasing size 
$N$)}
\end{table}

\begin{table}[htbp]
\brt
\begin{tabular}{cc}
N & \hspace{3cm}$p_{min}^{\rt}$\\
\hl
4 & \hspace{3cm}.104208\\
16 &\hspace{3cm} .750207\\
36 & \hspace{3cm} .840892
\end{tabular}
\ert
\caption{\label{pmintab2d} Values of $p_{min}^{\rt}$
(2D periodic square planar lattice of increasing (even-valued) size $N$)}
\end{table}

\begin{table}[htbp]
\brt
\begin{tabular}{ccc}
N & \hspace{3cm}$\Delta\, (1D)$ &\hspace{3cm}  $\Delta\, (2D)$\\
\hl
4 & \hspace{3cm}0.111 &\hspace{3cm} 0.802\\
16 &\hspace{3cm} 0.067 &\hspace{3cm} 0.416 \\
36 & \hspace{3cm} 0.016 &\hspace{3cm} 0.344
\end{tabular}
\ert
\caption{\label{compdelta} Relative enhancement $\Delta$ of the
global reaction time for 1D periodic lattices and 
2D periodic square planar lattices with the same number of sites $N$.}
\end{table}

\begin{table}[htbp]
\brt
\begin{tabular}{cc}
$N$ & $\langle v \rangle $ \\
\hl
4 & $-(2/3)(1240p+31p^3-560p^2-768)/(p^2+6p-8)^2$\\
&\\
9 & $-8(288p+2p^3-49p^2-584)/(p^2-8)^2$\\
&\\
16 & $-(8/135)(195677696p+15943296p^4-5833248p^3-115741696p^2-511812p^5$\\
& $-705816p^6+10719p^7-88932352)/(3p^4+24p^3-56p^2-96p+128)^2$\\
&\\
25 & $-(4/3)(16064512p-8024128p^4-3642368p^3+35447808p^2-2255p^8
-48480256$\\
& $-813472p^5+187628p^6+10448p^7+38p^9)/(p^5+4p^4-44p^3
-144p^2+64p+256)^2$\\
&\\
36 & $-(2/8575)(1690178540273664p-9697076228p^{13}+226056069312p^{12}
+17307675p^{15}$\\
& $+1052496354082816p^4-1589772806520832p^3-494222977269760p^2
+435557221152p^{11}$\\
& $+87781881765888p^8+477281389805568p^5-477098773250048p^6-39787096070144p^7$
\\
& $-7123662261248p^{10}-2654087139328p^9-697746159828992-2669903720p^{14})$\\
& $/(-6144p+2688p^4+5376p^3-8704p^2+5p^8+8192-1176p^5-256p^6+54p^7)^2$
\end{tabular}
\ert
\caption{\label{vp2d} Polynomial expressions for $\vrt$ 
(2D periodic square planar lattice of increasing size $N$)}
\end{table}

\begin{table}[htbp]
\brt
\begin{tabular}{cc}
N & \hspace{3cm}$p_{min}^{\vrt}$\\
\hl
4 & \hspace{3cm}.184684\\
16 & \hspace{3cm}.748812\\
36 & \hspace{3cm}.840409
\end{tabular}
\ert
\caption{\label{pminn2d}
Values of $p_{min}^{\vrt}$ 
(2D periodic square planar lattice of increasing (even-valued) size $N$)}
\end{table}

\section*{Figures}

\begin{appendix}

\section*{Appendix A: Tables with analytic results for the general case in 1D}

\begin{table}[htbp]
\brt
\begin{tabular}{cc}
$N$ & $\langle n \rangle $ \\
\hl
2 & $2/(2p_1+2p_2-3p_1 p_2)$\\
& \\
3 & $2/(p_1+p_2-p_1 p_2)$ \\
& \\
4 & $(10/3)(4p_1+4p_2-7p_1p_2)/(4p_1^2+8p_1p_2
-10p_1^2p_2+4p_2^2-10p_1p_2^2+5p_1^2p_2^2)$\\
& \\
5 &  $ (20 p_1 +20 p_2-28p_1p_2)/((p_1 p_2-2 p_1-2 p_2) 
(3p_1 p_2-2 p_1-2 p_2))$\\
& \\
6 &  $(28/5)(-20p_1p_2-10p_1^2+30p_1^2p_2-10p_2^2+30p_1p_2^2-21p_1^2p_2^2)
/(-24p_1^2p_2$\\
& $-24p_1p_2^2+56p_1^2p_2^2-8p_1^3-8p_2^3+
28p_1^3p_2-28p_1^3p_2^2+28p_1p_2^3-28p_1^2p_2^3+7p_1^3p_2^3)$\\
& \\
7 & $(-4/3)(21p_1^2p_2^2-36p_1^2p_2+14p_1^2-36p_1p_2^2+28p_1p_2+14p_2^2)/$\\
& $((p_1^2p_2^2-4p_1^2p_2+2p_1^2-4p_1p_2^2+4p_1p_2+2p_2^2)(-p_1-p_2+p_1p_2))$\\
& \\
8 & $(12/7)(-231p_1^3p_2^3+594p_1^3p_2^2-462p_1^3p_2+112p_1^3+
594p_1^2p_2^3-924p_1^2p_2^2$\\
& $+336p_1^2p_2-462p_1p_2^3+336p_1p_2^2+112p_2^3)/
((3p_1^3p_2^3-18p_1^3p_2^2+24p_1^3p_2-8p_1^3$\\
&$-18p_1^2p_2^3
+48p_1^2p_2^2-24p_1^2p_2+24p_1p_2^3-24p_1p_2^2-8p_2^3)(-2p_1-2p_2+3p_1p_2))$\\
& \\
9& 
$-10(33p_1^3p_2^3-99p_1^3p_2^2+88p_1^3p_2-24p_1^3-99p_1^2p_2^3+176p_1^2p_2^2
-72p_1^2p_2$\\
&$+88p_1p_2^3-72p_1p_2^2-24p_2^3)/((p_1^2p_2^2-6p_1^2p_2+4p_1^2
-6p_1p_2^2+8p_1p_2$\\
&$+4p_2^2)(4p_1^2+8p_1p_2
-10p_1^2p_2+4p_2^2-10p_1p_2^2+5p_1^2p_2^2))$\\
& \\
10&$(-22/9)(429p_1^4p_2^4-1716p_1^3p_2^4+240p_2^4+2288p_1^2p_2^4
-1248p_1p_2^4-1716p_1^4p_2^3-1248p_1^4p_2$\\
&$+240p_1^4+2288p_1^4p_2^2-3744p_1^2p_2^3+960p_1p_2^3
-3744p_1^3p_2^2+960p_1^3p_2
+1440p_1^2p_2^2+4576p_1^3p_2^3)/$\\
& $(11p_1^5p_2^5-110p_1^5p_2^4+308p_1^5p_2^3-352p_1^5p_2^2+176p_1^5p_2
-32p_1^5-110p_1^4p_2^5+616p_1^4p_2^4$\\
&$-1056p_1^4p_2^3+704p_1^4p_2^2-160p_1^4p_2+308p_1^3p_2^5-1056p_1^3p_2^4
+1056p_1^3p_2^3-320p_1^3p_2^2$\\
&$-352p_2^5p_1^2
+704p_1^2p_2^4-320p_1^2p_2^3+176p_1p_2^5-160p_1p_2^4-32p_2^5)$\\
\end{tabular}
\ert
\caption{\it \label{tabenct2} Polynomial expressions for 
$\langle n \rangle$ in the general case (1D periodic lattices of increasing
size $N$).}
\end{table}

\begin{table}[htbp]
\brt
\begin{tabular}{cc}
N & \hspace{3cm}$p_{1,c}^{\rt}$\\
\hl
4 & \hspace{3cm}.833972\\
6 & \hspace{3cm}.902207\\
8 & \hspace{3cm}.936378\\
10& \hspace{3cm}.955625\\
12& \hspace{3cm}.967419
\end{tabular}
\ert
\caption{\label{p1c1d} Critical values $p_{1,c}^{\rt}$ 
(1D periodic lattices of increasing size $N$).}
\end{table}

\begin{table}[htbp]
\brt
\begin{tabular}{cc}
$N$ & $\langle v \rangle $ \\
\hl
2 & $2(2-2p_1-2p_2+3p_1p_2)/(-2p_1-2p_2+3p_1p_2)^2$\\
& \\
3& $2(2-p_1-p_2+p_1p_2)/(-p_1-p_2+p_1p_2)^2$\\
& \\
4& 
$(2/3)(272p_1^2+544p_1p_2-1200p_1p_2^2+1530p_1^2p_2^2
+340p_1^3p_2-450p_1^3p_2^2$\\
&$+340p_2^3p_1-450p_1^2p_2^3+175p_1^3p_2^3-1200p_1^2p_2-80p_1^3-80p_2^3$\\
&$+272p_2^2)/(4p_1^2+8p_1p_2-10p_1^2p_2+4p_2^2-10p_1p_2^2+5p_1^2p_2^2)^2$\\
& \\
5& $4(21p_1^3p_2^3-71p_1^3p_2^2+68p_1^3p_2-20p_1^3-
71p_1^2p_2^3+348p_1^2p_2^2-356p_1^2p_2+
104p_1^2+68p_2^3p_1$\\
&$-356p_1p_2^2+208p_1p_2-20p_2^3+104p_2^2)/
((p_1p_2-2p_1-2p_2)^2(-2p_1-2p_2+3p_1p_2)^2)$\\
&\\
6& $
(28/5)(-3984p_1p_2^4+10056p_1^2p_2^4-80p_1^5+592p_2^4-
11592p_1^4p_2^3-1288p_1^5p_2^2$\\
&$+1498p_1^5p_2^3-11592p_1^3p_2^4+5754p_1^4p_2^4-798p_1^5p_2^4+520p_1^5p_2
-3984p_1^4p_2+10056p_1^4p_2^2$\\
&$+592p_1^4+3552p_1^2p_2^2+19072p_1^3p_2^3-11552p_1^3p_2^2+2368p_2^3p_1
-11552p_1^2p_2^3+2368p_1^3p_2$\\
&$+1498p_1^3p_2^5-798p_1^4p_2^5+147p_1^5p_2^5-80p_2^5+520p_2^5p_1
-1288p_2^5p_1^2)/(-24p_1^2p_2-24p_1p_2^2$\\
&$+56p_1^2p_2^2+7p_1^3p_2^3-28p_1^3p_2^2+28p_2^3p_1-28p_1^2p_2^3+28p_1^3p_2
-8p_1^3-8p_2^3)^2$\\
&\\
7&$
(4/3)(-1596p_1p_2^4+3384p_1^2p_2^4-28p_1^5+280p_2^4-3240p_1^4p_2^3
-328p_1^5p_2^2+320p_1^5p_2^3$\\
&$-3240p_1^3p_2^4+1314p_1^4p_2^4-141p_1^5p_2^4+156p_1^5p_2-1596p_1^4p_2
+3384p_1^4p_2^2+280p_1^4$\\
&$+1680p_1^2p_2^2+6456p_1^3p_2^3-4648p_1^3p_2^2+1120p_2^3p_1-4648p_1^2p_2^3
+1120p_1^3p_2+320p_1^3p_2^5$\\
&$-141p_1^4p_2^5+21p_1^5p_2^5-28p_2^5+156p_2^5p_1-328p_2^5p_1^2)
/((p_1^2p_2^2-4p_1^2p_2+2p_1^2$\\
&$-4p_1p_2^2+4p_1p_2+2p_2^2)^2(-p_1-p_2+p_1p_2)^2)$\\
&\\
\end{tabular}
\ert
\caption{\it \label{tabenct3} Polynomial expressions for 
$\vrt$ in the general case (1D periodic lattices of size $N=2$ to $N=7$).}
\end{table}

\begin{table}[htbp]
\brt
\begin{tabular}{cc}
$N$ & $\langle v \rangle $ \\
\hl
8&$(12/7)(349440p_1^2p_2^4-1998080p_1^4p_2^3-1005312p_1^5p_2^2+
2845920p_1^5p_2^3-1998080p_1^3p_2^4$\\
&$+4230240p_1^4p_2^4
-3982752p_1^5p_2^4+139776p_1^5p_2+349440p_1^4p_2^2+465920p_1^3p_2^3
-54864p_2^7p_1^2$\\
&$-54864p_1^7p_2^2
+64746p_1^5p_2^7+103080p_1^3p_2^7-19206p_1^7p_2^6-19206p_1^6p_2^7
+2079p_1^7p_2^7$\\
&$-109512p_1^7p_2^4
+64746p_1^7p_2^5+103080p_1^7p_2^3-109512p_1^4p_2^7+236610p_1^6p_2^6-
867528p_1^5p_2^6$\\
&$+746256p_2^6p_1^2
-1419024p_2^6p_1^3+1505928p_2^6p_1^4-867528p_1^6p_2^5-1419024p_1^6p_2^3
+1505928p_1^6p_2^4$\\
&$+746256p_1^6p_2^2+23296p_2^6+15456p_1^7p_2+15456p_2^7p_1-1792p_1^7+23296p_1^6-206080p_1^6p_2
$\\
&$-206080p_2^6p_1+2845920p_1^3p_2^5-3982752p_1^4p_2^5+2805696p_1^5p_2^5
+139776p_2^5p_1$\\
&$-1005312p_2^5p_1^2
-1792p_2^7)/((3p_1^3p_2^3-18p_1^3p_2^2+24p_1^3p_2-8p_1^3-18p_1^2p_2^3
+48p_1^2p_2^2$\\
&$-24p_1^2p_2+24p_2^3p_1-24p_1p_2^2-8p_2^3)^2(-2p_1-2p_2+3p_1p_2)^2)$\\
& \\
9& $2(3000448p_1^5p_2^3-1203840p_1^5p_2^2+188928p_1^5p_2-3669728p_1^5p_2^4
+3000448p_1^3p_2^5-1203840p_1^2p_2^5$\\
&$+188928p_1p_2^5-3669728p_1^4p_2^5
+2238576p_1^5p_2^5+629760p_1^3p_2^3+472320p_1^4p_2^2-2394240p_1^4p_2^3
$\\
&$+472320p_1^2p_2^4-2394240p_1^3p_2^4+4463872p_1^4p_2^4+825p_1^7p_2^7
+783232p_1^6p_2^2-1300176p_1^6p_2^3$\\
&$+1195368p_1^6p_2^4-591060p_1^6p_2^5
-70340p_1^4p_2^7+76080p_1^3p_2^7-46160p_1^2p_2^7+35860p_1^5p_2^7$\\
&$-9075p_1^6p_2^7-70340p_1^7p_2^4+76080p_1^7p_2^3-46160p_1^7p_2^2
+35860p_1^7p_2^5-591060p_1^5p_2^6+1195368p_1^4p_2^6$\\
&$-1300176p_1^3p_2^6
+136620p_1^6p_2^6-9075p_1^7p_2^6+14720p_1^7p_2+783232p_1^2p_2^6
+14720p_1p_2^7+31488p_2^6$\\
&$+31488p_1^6-1920p_1^7-1920p_2^7-246144p_1^6p_2
-246144p_1p_2^6)/((p_1^2p_2^2-6p_1p_2^2+4p_2^2-6p_1^2p_2$\\
&$+8p_1p_2+4p_1^2)^2(4p_2^2+8p_1p_2-10p_1p_2^2+4p_1^2-10p_1^2p_2
+5p_1^2p_2^2)^2)$\\
& \\
10&$(22/9)(8687616p_1^5p_2^3-57738240p_1^5p_2^4+8687616p_1^3p_2^5
-57738240p_1^4p_2^5+150474240p_1^5p_2^5$\\
&$-97613824p_1^7p_2^4+10859520p_1^4p_2^4+1236378p_1^8p_2^8+37019488p_1^7p_2^7
+20033904p_1^6p_2^8$\\
&$-7090512p_1^7p_2^8
+155136p_2^8-20579328p_1^8p_2^3+32759936p_1^8p_2^4-32651520p_1^8p_2^5
-1691136p_1^8p_2$\\
&$+20033904p_1^8p_2^6-7090512p_1^8p_2^7
+7893504p_1^2p_2^8-20579328p_1^3p_2^8+32759936p_1^4p_2^8-32651520p_1^5p_2^8$\\
&$-1691136p_1p_2^8+7893504p_1^8p_2^2+4343808p_1^6p_2^2-34707456p_1^6p_2^3+113143296p_1^6p_2^4
-192586240p_1^6p_2^5$\\
&$+155136p_1^8
-97613824p_1^4p_2^7+45717504p_1^3p_2^7-11630592p_1^2p_2^7+122302400p_1^5p_2^7-90295040p_1^6p_2^7$\\
&$+45717504p_1^7p_2^3-11630592p_1^7p_2^2
+122302400p_1^7p_2^5-192586240p_1^5p_2^6+113143296p_1^4p_2^6$\\
&$-34707456p_1^3p_2^6+181026560p_1^6p_2^6-90295040p_1^7p_2^6
+1241088p_1^7p_2+1241088p_1p_2^7+4343808p_1^2p_2^6$\\
&$-7680p_1^9-7680p_2^9-1531904p_1^9p_2^4
+1524160p_1^9p_2^5-944944p_1^9p_2^6+346060p_1^9p_2^7+970816p_1^9p_2^3$\\
&$-377344p_1^9p_2^2+82176p_1^9p_2
-66066p_1^9p_2^8-1531904p_1^4p_2^9+1524160p_1^5p_2^9-944944p_1^6p_2^9$\\
&$+346060p_1^7p_2^9+970816p_1^3p_2^9
-377344p_1^2p_2^9+82176p_1p_2^9-66066p_1^8p_2^9+4719p_1^9p_2^9)/
(11p_1^5p_2^5$\\
&$-110p_1^4p_2^5+308p_1^3p_2^5-352p_1^2p_2^5+176p_1p_2^5-32p_2^5-110p_1^5p_2^4
+616p_1^4p_2^4-1056p_1^3p_2^4$\\
&$+704p_1^2p_2^4-160p_1p_2^4+308p_1^5p_2^3-1056p_1^4p_2^3+1056p_1^3p_2^3
-320p_1^2p_2^3-352p_1^5p_2^2$\\
&$+704p_1^4p_2^2-320p_1^3p_2^2+176p_1^5p_2-160p_1^4p_2-32p_1^5)^2$
\end{tabular}
\ert
\caption{\it \label{tabenct5} Polynomial expressions for 
$\langle v \rangle$ in the general case
(1D periodic lattices of size $N=7$ to $N=10$).}
\end{table}

\begin{table}[htbp]
\brt
\begin{tabular}{cc}
$N$ &\hspace{3cm} $p_{1,c}^{\vrt}$\\
\hl
4 & \hspace{3cm} .817442 \\
6 & \hspace{3cm} .891104 \\
8 & \hspace{3cm}.929001 \\
10 &\hspace{3cm} .950475 \\
\end{tabular}
\ert
\caption{\it \label{p1c1e} 
Critical values $p_{1,c}^{\vrt}$ (1D periodic lattices of increasing
size).}
\end{table}

\section*{Appendix B: Tables with analytic results for the general case in 2D}

\begin{table}[htbp]
\brt
\begin{tabular}{cc}
$N$ & $\langle n \rangle $ \\
\hl
4 & $-(2/3)(71p_1p_2-40p_2-40p_1)/(13p_1^2p_2^2-22p_1p_2^2-22p_1^2p_2
+8p_2^2+16p_1p_2+8p_1^2)$\\
& \\
9 & $-8(11p_1p_2-9p_1-9p_2)/(7p_1^2p_2^2-16p_1^2p_2-16p_1p_2^2+8p_1^2
+16p_1p_2+8p_2^2)$\\
&\\
16 & $-(8/45)(26497p_1^3p_2^3-68308p_1^3p_2^2+53804p_1^3p_2-13184p_1^3
-68308p_1^2p_2^3+107608p_1^2p_2^2$\\
&$-39552p_1^2p_2+53804p_1p_2^3
-39552p_1p_2^2-13184p_2^3)/(147p_1^4p_2^4-664p_1^4p_2^3
+1000p_1^4p_2^2$\\
&$-608p_1^4p_2+128p_1^4-664p_1^3p_2^4+2000p_1^3p_2^3
-1824p_1^3p_2^2+512p_1^3p_2+1000p_1^2p_2^4$\\
&$-1824p_1^2p_2^3+768p_1^2p_2^2-608p_1p_2^4+512p_1p_2^3+128p_2^4)$\\
&\\
$25$ & $-(76/3)(1920p_1^2p_2^2-3984p_1^3p_2^2+1280p_1^3p_2-1185p_1^3p_2^4
-1185p_1^4p_2^3$\\
& $+1940p_1^4p_2^2+1940p_1^2p_2^4+1280p_1p_2^3+3880p_1^3p_2^3-3984p_1^2p_2^3
-1328p_1^4p_2$\\
&$+320p_2^4+255p_1^4p_2^4-1328p_2^4p_1+320p_1^4)/(95p_1^5p_2^5-684p_1^5p_2^4
-2560p_1^3p_2^2$\\
&$+1788p_1^5p_2^3-2160p_1^5p_2^2+1216p_1^5p_2-256p_1^5-684p_1^4p_2^5
+3576p_1^4p_2^4$\\
&$-6480p_1^4p_2^3+4864p_1^4p_2^2-1280p_1^4p_2+1788p_1^3p_2^5-6480p_1^3p_2^4+7296p_1^3p_2^3$\\
&$-2160p_1^2p_2^5+4864p_1^2p_2^4-2560p_1^2p_2^3
+1216p_1p_2^5-1280p_2^4p_1-256p_2^5)$\\
& \\
36 & 
$-(2/1225)(-12930686208p_1^7p_2^4-12930686208p_1^4p_2^7+15328776360p_1^6p_2^6$\\
&$-38792058624p_1^5p_2^6-5200699392p_1^5p_2^2-5200699392p_2^5p_1^2
+74710699776p_1^5p_2^5$\\
&$+30453427200p_1^5p_2^3-68813798400p_1^4p_2^5+30453427200p_1^3p_2^5
+49807133184p_1^6p_2^4$\\
&$-38792058624p_1^6p_2^5-34406899200p_1^6p_2^3+49807133184p_1^4p_2^6-34406899200p_1^3p_2^6$\\
&$+12181370880p_1^2p_2^6-68813798400p_1^5p_2^4-1733566464p_1^6p_2-1733566464p_1p_2^6$\\
&$+12181370880p_1^6p_2^2+40604569600p_1^4p_2^4
-8667832320p_1^3p_2^4-8667832320p_1^4p_2^3$\\
&$+298340359p_1^7p_2^7-2385516420p_1^7p_2^6-2385516420p_1^6p_2^7+7664388180p_1^7p_2^5+7664388180p_1^5p_2^7$\\
&$-247652352p_2^7+12451783296p_1^3p_2^7-6881379840p_1^2p_2^7+2030228480p_1p_2^7-247652352p_1^7$\\
&$+12451783296p_1^7p_2^3-6881379840p_1^7p_2^2+2030228480p_1^7p_2)/(193480p_2^8p_1^6-71680p_1p_2^8$\\
&$-530176p_1^3p_2^8+263680p_1^2p_2^8-457880p_1^5p_2^8+633728p_1^4p_2^8-71680p_1^8p_2-2650880p_1^7p_2^4$\\
&$-457880p_1^8p_2^5-530176p_1^8p_2^3+263680p_1^8p_2^2+633728p_1^8p_2^4-2650880p_1^4p_2^7+3802368p_1^6p_2^6$\\
&$-5301760p_1^6p_2^5-5301760p_1^5p_2^6+5273600p_1^5p_2^5-2508800p_1^5p_2^4+458752p_1^5p_2^3$\\
&$-2508800p_1^4p_2^5+458752p_1^3p_2^5+3955200p_1^6p_2^4-1505280p_1^6p_2^3+3955200p_1^4p_2^6$\\
&$-1505280p_1^3p_2^6+229376p_1^2p_2^6+229376p_1^6p_2^2+573440p_1^4p_2^4+193480p_1^8p_2^6+8192p_1^8$\\
&$+386960p_1^7p_2^7-1373640p_1^7p_2^6-1373640p_1^6p_2^7+2534912p_1^7p_2^5+2534912p_1^5p_2^7$\\
&$+3815p_1^8p_2^8-43154p_1^8p_2^7-43154p_1^7p_2^8+8192p_2^8+1582080p_1^3p_2^7-501760p_1^2p_2^7$\\
&$+65536p_1p_2^7+1582080p_1^7p_2^3-501760p_1^7p_2^2+65536p_1^7p_2)$
\end{tabular}
\ert
\caption{\label{np1p22d} Polynomial expressions 
for $\langle n\rangle$ in the general case  
(2D periodic square planar lattices of increasing size)}
\end{table}

\begin{table}[htbp]
\brt
\begin{tabular}{cc}
N &\hspace{3cm} $p_{1,c}^{\rt}$\\
\hl
4 &\hspace{3cm} .710835\\
16 &\hspace{3cm} .842153\\
36 &\hspace{3cm} .887094
\end{tabular}
\ert
\caption{\label{p12d} Critical values $p_{1,c}^{\rt}$ (2D square 
planar lattices of increasing size with an even value of $N$).}
\end{table}

\begin{table}[htbp]
\brt
\begin{tabular}{cc}
$N$ & $\langle v \rangle $ \\
\hl
4 & $(2/3)
(2176p_1p_2-2082p_1^2p_2^3+923p_1^3p_2^3-2082p_1^3p_2^2+1448p_2^3p_1$
\\
& $+1448p_1^3p_2-320p_1^3-320p_2^3+1088p_1^2-4864p_1^2p_2+1088p_2^2
-4864p_1p_2^2$\\
& $+6418p_1^2p_2^2)
/(13p_1^2p_2^2-22p_1p_2^2+8p_2^2-22p_1^2p_2+16p_1p_2+8p_1^2)^2$\\
&\\
9& 
$8(1312p_1p_2-239p_1^2p_2^3+77p_1^3p_2^3-239p_1^3p_2^2+232p_2^3p_1
+232p_1^3p_2$\\
&$-72p_1^3-72p_2^3+656p_1^2-1832p_1^2p_2+656p_2^2-1832p_1p_2^2
+1464p_1^2p_2^2)/$\\
&$(-16p_1^2p_2-16p_1p_2^2+7p_1^2p_2^2+8p_1^2
+16p_1p_2+8p_2^2)^2$\\
&\\
16 & $(8/135)(44708352p_1^7p_2+44708352p_2^7p_1+1879900160p_1^3p_2^3
+1409925120p_1^4p_2^2$\\
&$-7866368000p_1^4p_2^3+1409925120p_2^4p_1^2-7866368000p_1^3p_2^4+16208510208p_1^4p_2^4
-3950903296p_1^2p_2^5$\\
&$-5062656p_1^7+10880187392p_1^3p_2^5-14811384576p_1^4p_2^5+10880187392p_1^5p_2^3-14811384576p_1^5p_2^4$\\
&$+10174354048p_1^5p_2^5+563970048p_2^5p_1+563970048p_1^5p_2+93995008p_1^6+93995008p_2^6$\\
&$-5062656p_2^7-3094999904p_1^5p_2^6+5409620192p_1^4p_2^6
-5210329472p_1^3p_2^6+2820640640p_1^2p_2^6$\\
&$+869041300p_1^6p_2^6-3094999904p_1^6p_2^5+5409620192p_1^6p_2^4
-5210329472p_1^6p_2^3+2820640640p_1^6p_2^2$\\
&$-804356096p_2^6p_1-804356096p_1^6p_2-366246240p_1^4p_2^7
+239288100p_1^5p_2^7-82905852p_1^6p_2^7$\\
&$-82905852p_1^7p_2^6+11685177p_1^7p_2^7+239288100p_1^7p_2^5
-366246240p_1^7p_2^4+322443168p_2^7p_1^3$\\
&$-3950903296p_1^5p_2^2+322443168p_2^3p_1^7-163920768p_1^7p_2^2-163920768p_2^7p_1^2)
/(147p_1^4p_2^4$\\
&$-664p_1^3p_2^4+1000p_2^4p_1^2-608p_2^4p_1+128p_2^4-664p_1^4p_2^3
+2000p_1^3p_2^3-1824p_1^2p_2^3$\\
&$+512p_2^3p_1+1000p_1^4p_2^2-1824p_1^3p_2^2+768p_1^2p_2^2-608p_1^4p_2+512p_1^3p_2+128p_1^4)^2$
\end{tabular}
\ert
\caption{\label{vp1p22d} 
Polynomial expressions for $\vrt$ in the general case 
(2D periodic square planar lattices of increasing size)}
\end{table}

\begin{table}[htbp]
\brt
\begin{tabular}{cc}
N & \hspace{3cm}$p_{1,c}^{\vrt}$\\
\hl
4 & \hspace{3cm}.702484\\
16 & \hspace{3cm}.837365\\
36 &\hspace{3cm}.885175
\end{tabular}
\ert
\caption{\label{pminv2d}
Critical values $p_{1,c}^{\vrt}$ (2D periodic square planar lattices
of increasing size with an even number of sites $N$).}
\end{table}

\end{appendix}

\end{document}